\documentclass[apj,twocolumn]{openjournal}

\usepackage{amsmath}
\usepackage{array}
\usepackage{booktabs}
\usepackage{multirow}
\usepackage{color}
\usepackage{soul}
\usepackage{capt-of}

\usepackage[dvipsnames]{xcolor} 

\usepackage[breaklinks,colorlinks,citecolor=blue,urlcolor=blue]{hyperref}

\newcommand{\red}[1]{\textcolor{red}{#1}}
\newcommand{\green}[1]{\textcolor{dkgreen}{#1}}

\defcitealias{El-Badry2023}{E23}
\usepackage{xspace}
\usepackage{orcidlink}

\newcommand{\Dsix}{\ensuremath{\mathrm{D}^{6}}\xspace}

\renewcommand{\arraystretch}{1.2}  

\usepackage{listings}
\usepackage{color}
\definecolor{dkgreen}{rgb}{0,0.6,0}
\definecolor{gray}{rgb}{0.5,0.5,0.5}
\definecolor{mauve}{rgb}{0.58,0,0.82}
\definecolor{golden}{rgb}{0.86,0.65,0.01}
\begin{document}


\title{A systematic survey for hypervelocity runaways from thermonuclear supernovae}

\author{\vspace{-1cm}Kareem El-Badry\,\orcidlink{0000-0002-6871-1752}$^{1}$}
\author{Klaus Werner\,\orcidlink{0000-0002-6428-2276}$^{2}$}
\author{Ken J. Shen\,\orcidlink{0000-0002-9632-6106}$^{3}$}
\author{Jay Strader\,\orcidlink{0000-0002-1468-9668}$^{4}$}
\author{Antonio C. Rodriguez\,\orcidlink{0000-0003-4189-9668}$^{5}$}
\author{Jiwon Jesse Han\,\orcidlink{0000-0002-6800-5778}$^{6}$}
\author{Vedant Chandra\,\orcidlink{0000-0002-0572-8012}$^{5}$}
\author{Laura Chomiuk\,\orcidlink{0000-0002-8400-3705}$^{4}$}
\author{Zachary P. Vanderbosch\,\orcidlink{0000-0002-0853-3464}$^{1}$}
\author{Lisa Blomberg\,\orcidlink{0009-0004-4475-2121}$^{1}$}
\author{Natsuko Yamaguchi\,\orcidlink{0000-0001-6970-1014}$^{1}$}
\author{Pranav Nagarajan\,\orcidlink{0000-0002-1386-0603}$^{1}$}
\author{Ilaria Caiazzo\,\orcidlink{0000-0002-4770-5388}$^{7}$}
\author{Jan van Roestel\,\orcidlink{0000-0002-2626-2872}$^{7}$}
\author{Hila Glanz\,\orcidlink{0000-0002-6012-2136}$^{1}$}
\author{Tin Long Sunny Wong\,\orcidlink{0000-0001-9195-7390}$^{8}$}
\author{Aakash Bhat\,\orcidlink{0000-0002-4803-5902}$^{9}$}
\author{Mark A. Hollands\,\orcidlink{0000-0003-0089-2080}$^{10}$}
\author{Boris T. G\"ansicke\,\orcidlink{0000-0002-2761-3005}$^{10}$}

\affiliation{$^1$Department of Astronomy, California Institute of Technology, 1200 E. California Blvd., Pasadena, CA 91125, USA}
\affiliation{$^2$Institute for Astronomy and Astrophysics, University of T\"ubingen, Sand 1, 72076 T\"ubingen, Germany}
\affiliation{$^3$Department of Astronomy and Theoretical Astrophysics Center, University of California, Berkeley, CA 94720, USA}
\affiliation{$^4$Center for Data Intensive and Time Domain Astronomy, Department of Physics and Astronomy, Michigan State University, East Lansing, MI 48824, USA}
\affiliation{$^5$Center for Astrophysics $|$ Harvard \& Smithsonian, 60 Garden Street, Cambridge, MA 02138, USA}
\affiliation{$^6$Kavli Institute for Particle Astrophysics and Cosmology, Stanford University, Stanford, CA 94305, USA}
\affiliation{$^7$Institute of Science and Technology Austria (ISTA), Am Campus 1, 3400 Klosterneuburg, Austria}
\affiliation{$^8$The Observatories of the Carnegie Institution for Science, 813 Santa Barbara Street, Pasadena, CA 91101, USA}
\affiliation{$^9$Institute of Physics and Astronomy, University of Potsdam, Karl-Liebknecht-Str. 24/25, 14476 Potsdam, Germany}
\affiliation{$^{10}$Department of Physics, University of Warwick, Coventry CV4 7AL, UK}
\email{Corresponding author: kelbadry@caltech.edu}

\begin{abstract}
The explosion of a white dwarf (WD) in a close binary can launch a surviving runaway star at velocities of $\gtrsim 1000\, \rm km\,s^{-1}$. Such runaways provide a direct probe of thermonuclear supernovae (SNe) in double-degenerate binaries. Several candidate runaways are known, but their evolutionary states and the demographics of the broader population are uncertain. To enable robust population inference, we carry out a systematic survey for hypervelocity runaways with a simple selection function, selecting candidates based on large {\it Gaia}-inferred tangential velocities and blue colors. We classify 100\% of the resulting 92 candidates using a combination of spectroscopic follow-up and archival data. 
The search yields ten suspected \Dsix stars and three LP 40-365 stars. Three \Dsix stars are new discoveries, including two hot ($T_{\rm eff}\gtrsim 50{,}000\,{\rm K}$) objects and one cool ($T_{\rm eff}\approx 7{,}000\,{\rm K}$) object. We forward-model our survey under several proposed \Dsix star evolutionary models, coupling each to a Galactic model and the survey selection function. No single model reproduces the observed diversity of \Dsix stars, which likely reflects a range of remnant masses, ages, and heating mechanisms. Models in which runaway companions are heated by SN shocks alone are too faint and short-lived to explain most of the observed sample, while fully reheated models are too luminous and long-lived. Models with intermediate heating, as occurs in some simulations of violent mergers and partially disrupted remnants, best match the observed magnitude, distance, and kinematic-age distributions. The inferred \Dsix star birth rate is model dependent, but the models that best match the observed population require rates of only a few percent of the Galactic SN Ia rate, perhaps implying that most SNe Ia result from WD binaries in which both components explode. If most SNe Ia do produce surviving runaways, these must be fainter or shorter-lived than the currently known runaways.

\end{abstract}

\maketitle

\section{Introduction}
\label{sec:intro}

Several classes of progenitor models for Type Ia supernovae (SNe Ia) and related thermonuclear transients begin with the gravitational wave-driven inspiral of two white dwarfs (WDs) in a close binary \citep[e.g.][]{Webbink1984, Iben1984}. In some models, only one of the WDs explodes. The other -- suddenly free of its companion's gravity -- is predicted to escape with a velocity comparable to its orbital velocity, which is typically $(1000-2000)\,{\rm km\,s^{-1}}$.

Over the last decade, searches using wide-field photometric and spectroscopic surveys have identified more than a dozen high-velocity stars suspected to have been ejected from thermonuclear explosions in WD binaries \citep{Geier2015, Vennes2017, Shen2018, Raddi2019,  El-Badry2023, Hollands2025, Bhat2026}. Most of these stars fall between the main sequence and WD cooling track in the color-magnitude diagram (CMD). The objects with space velocities $\gtrsim 1000\,{\rm km\,s^{-1}}$ have been dubbed ``\Dsix stars'', after a theoretical model for their formation \citep[``dynamically driven, double-degenerate, double-detonation'' SNe;][]{Shen2018b}. Some recent works have proposed progenitor models for these objects that are distinct from the \Dsix scenario \citep{Pakmor2025, Bhat2025, Glanz2025}, in which a violent WD merger leads to detonation of the primary and partial disruption of the secondary. In this case, the hypervelocity star can reach lower masses and somewhat higher velocities than in the \Dsix scenario. We refer to the observed runaway stars with velocities $\gtrsim 1000\,{\rm km\,s^{-1}}$ as \Dsix stars in this work for consistency with the previous literature, but we are agnostic of the precise mechanism that detonated the SNe from which they were ejected.

The observed sample of thermonuclear-runaway candidates also includes LP~40-365 stars \citep{Vennes2017, Raddi2019, Bhat2026}, of which eight are currently known (Section~\ref{sec:all_known}), and runaway helium stars, of which one is known \citep[e.g.][]{Geier2015, Neunteufel2022, Geier2024}. LP~40-365 stars have lower inferred ejection velocities, typically $\sim 600\,{\rm km\,s^{-1}}$, and neon + oxygen dominated atmospheres polluted with partially burned SN ash. They have been associated with failed or peculiar thermonuclear explosions, such as SNe Iax \citep{Foley2013}, in which the observed star is the remnant of a partially burned ONe WD rather than the surviving donor. While both LP~40-365 stars and \Dsix stars are likely products of thermonuclear SNe in close binaries containing a WD, they likely trace different outcomes: partially burned primaries for LP~40-365 stars, and surviving donors or donor fragments for \Dsix stars. Like \Dsix stars, runaway helium stars are likely surviving donors, but they are ejected from WD + helium star binaries rather than WD + WD binaries and thus have lower velocities.

The eight suspected \Dsix stars published to date have radii of $(0.02-0.2)\,R_{\odot}$, significantly larger than ordinary WDs. They span a wide range of temperatures, from $\sim 7000$\,K to $\sim 100,000$\,K \citep{Chandra2022, Werner2024, Hollands2025}.
Seven have hydrogen-free atmospheres, likely because their envelopes were stripped by mass transfer prior to their companions' detonation.  Most appear to have atmospheres dominated by carbon and oxygen, together with  enhancements in metals such as iron, nickel, and silicon \citep[]{El-Badry2023, Werner2024, Werner2024a, Hollands2025}. At least one has a helium-rich surface \citep{Werner2024}; the same object has hydrogen detected in its atmosphere, the origin of which is uncertain.

The masses of the known \Dsix stars are in most cases poorly constrained. Their inferred ejection velocities span $(1000-3000)\,{\rm km\,s^{-1}}$. It is possible to infer the stars' masses from their velocities if they were semi-detached donors that were launched from Keplerian pre-SN orbits \citep{Bauer2021, Braudo2024}, but such constraints could be biased if the stars were partially disrupted during or just prior to their ejection \citep[e.g.][]{Wong2025, Bhat2025, Pakmor2025}. Masses can in principle be measured directly from constraints on the stars' radii and surface gravities \citep[e.g.][]{Hollands2025}, but uncertainties in atmospheric models in most cases limit the robustness of such constraints. 

One cool \Dsix star is on a trajectory that traces back to a SN Ia remnant \citep{Shen2018, Chandra2022}. Assuming this association is physical and not a chance alignment, the object's kinematic age is $\sim 0.1$ Myr. Tracing the trajectories of the other stars back to the Galactic plane suggests typical kinematic ages of $\sim 1$ Myr, though the stars' uncertain birth locations lead to uncertainty in this value \citep{El-Badry2023}. The evolutionary state of \Dsix stars is uncertain. Several models have been explored to explain their inflated radii and wide range of temperatures \citep[][]{Zhang2019, Bauer2019, Bhat2025b, Yamaguchi2025, Wong2024, Wong2025, Bhat2025, Shen2025}. While some models have succeeded in matching the observed temperatures, luminosities, velocities, and ages of individual \Dsix stars, none reproduce the diversity of physical properties in the observed population. For the cool \Dsix stars ($T_{\rm eff}\lesssim 10,000\,{\rm K}$), the models that currently appear most promising are those in which the donors lost significant mass during or just prior to their companions' explosions \citep{Shen2025, Wong2025, Bhat2025}, such that the stars have lower masses than would be inferred when their ejection velocities are interpreted as pre-explosion orbital velocities. The hotter objects appear to be more massive than the cool objects. Most models have struggled to explain why the hot objects remain inflated at relatively old ages.

If most SNe Ia arise from WD+WD mergers in which one component survives, then the birth rate of \Dsix stars should be similar to the SNe Ia rate \citep[$\mathcal{R}_{\rm Ia}\approx 0.005\,\rm yr^{-1}$ in the Milky Way;][]{Li2011, Maoz2014}. On the other hand, if most SNe Ia arise from other channels -- such as WD+WD mergers in which both stars explode \citep[e.g.][]{Pakmor2022, Boos2024, Pollin2024, Prust2026, Mehta2026}, or binaries containing a WD and a nondegenerate star \citep[e.g.][]{Whelan1973} -- the birth rate of \Dsix stars should be much lower. Observationally constraining the  birth rate of \Dsix stars is thus a promising avenue to constrain the dominant progenitor channel for SNe Ia. 

In this paper, we present results of a systematic search for \Dsix stars using astrometry from {\it Gaia} and spectroscopic follow-up. Our search is designed to have a simple selection function, making it amenable to population modeling. Early discoveries from the search were presented by \citet[][hereafter E23]{El-Badry2023}; here, we present the complete survey and several new discoveries. We also compare the survey's yield to several classes of evolutionary models for SN survivors developed in the last few years.

The remainder of this paper is organized as follows. Section~\ref{sec:candidates} describes how we select candidate \Dsix stars based on {\it Gaia} astrometry and photometry. Section~\ref{sec:spectra} details our spectroscopic follow-up, while Section~\ref{sec:discoveries} presents our newly discovered \Dsix star candidates. We integrate the stars' orbits in Section~\ref{sec:orbits} and compare their demographics to simulated populations in Section~\ref{sec:space_density}. We summarize our main results in Section~\ref{sec:conclusions}.

\section{Candidate selection}
\label{sec:candidates}
Following \citetalias{El-Badry2023}, we  select candidate hypervelocity runaways using cuts on astrometric tangential velocity, proper motion, color, and apparent magnitude.

\subsection{Tangential velocity and proper motion}
\label{sec:vtan}

Given a total proper motion $\mu$ and parallax $\varpi$, the implied tangential velocity is 
\begin{equation}
    \label{vperp}
    v_{\perp}=4.74\,{\rm km\,s^{-1}}\times\left(\frac{\mu}{{\rm mas}\,{\rm yr}^{-1}}\right)\left(\frac{\varpi}{{\rm mas}}\right)^{-1}.
\end{equation} 
We select sources with $v_{\perp} > 600\,\rm km\,s^{-1}$. This threshold is large enough to exclude most bound halo stars while including a majority of \Dsix stars. Sources on near-radial trajectories will be missed by this selection; blind spectroscopic searches may discover them in the future.

For the sources of interest to our search, the uncertainty in $v_{\perp}$ is always dominated by parallax uncertainty. Many distant sources with small, noise-dominated parallaxes and relatively low proper motions satisfy $v_{\perp} > 600\,\rm km\,s^{-1}$ because parallax inversion overestimates their distances. We reduce contamination from such sources using two cuts. First, we require a minimum proper motion, $\mu > 50\,{\rm mas\,yr^{-1}}$, which effectively limits the search to sources with distances $d \lesssim 4.22\,{\rm kpc}\times\left[v_{\perp,\,{\rm true}}/\left(1000\,{\rm km\,s^{-1}}\right)\right]$ for a true tangential velocity $v_{\rm \perp,\, true}$. Second, we define 
$v_{\perp,{\rm lower}}=4.74\mu/\left(\varpi+\sigma_{\varpi}\right)$ as a rough proxy for the $1\sigma$ lower limit on $v_\perp$, and we require $v_{\perp,{\rm lower}}> 400\,{\rm km\,s^{-1}}$. Here $\sigma_{\varpi}$ is the parallax uncertainty. 

Reducing the thresholds of $v_{\perp}$, $v_{\perp,\,{\rm lower}}$, and $\mu$ increases the number of candidates, improving the search's sensitivity to sources at larger distances and sources on more radial trajectories. However, it also dramatically increases the number of contaminants.  We have found that the adopted thresholds provide a reasonable compromise between completeness and purity. The adopted thresholds are accounted for in our modeling (Section~\ref{sec:space_density}), so their precise values are ultimately unimportant as long as spectroscopic follow-up can reliably distinguish between \Dsix stars and contaminants.  

In addition to the above cuts, we require \texttt{ruwe} $< 1.4$, since sources with larger \texttt{ruwe} generally have problematic astrometric solutions affected by binarity or blending with nearby sources \citep{Lindegren_2018_RUWE, Belokurov2020, El-Badry2025}.  This cut is expected to exclude only a small fraction of true \Dsix stars, which should not be in binaries and are usually detected in regions that are not particularly crowded. 

\subsection{Color}
All currently known \Dsix stars -- as well as related classes of SN survivors such as LP 40-365 stars -- have relatively blue intrinsic colors. To reduce contamination from main-sequence halo stars, we require $G_{\rm BP} - G_{\rm RP} < 0.5$. This dramatically reduces the number of contaminants: removing it increases the number of candidates by a factor of 400. The price to pay is that our search is not sensitive to red objects, including both intrinsically red sources with $T_{\rm eff} \lesssim 6500\,{\rm K}$ \citep[some of which have been proposed to exist; e.g.][]{Shen2025, Wong2025} and intrinsically blue sources with significant foreground extinction. 

We explored selections based on extinction-corrected color, which could in principle make our search sensitive to more sources in the Galactic plane. Because most of our candidates have noisy parallaxes, we found that obtaining robust extinctions for all of them using 3D dust maps is currently impractical. 
Moreover, since most known \Dsix stars are already near the faint magnitude limit of our search (Section~\ref{sec:mag_cut}) and extinction makes sources both redder and fainter, our simulations in Section~\ref{sec:space_density} predict that the number of otherwise promising \Dsix stars our search misses due to extinction is modest. Thus, our selection is based on observed color without extinction corrections.

\subsection{Apparent magnitude}
\label{sec:mag_cut}
We adopt a magnitude limit of $G <  20.5$, just above the  {\it Gaia} faint limit of $G\simeq 20.7$. Although only 23\% of sources in {\it Gaia} DR3 have $G>20.5$, removing this faint limit increases the number of candidates by a factor of 2.8. This reflects the fact that sources near the faint limit have the largest parallax uncertainties and are most likely to have spurious colors or astrometry.

\subsection{ADQL query}
\label{sec:adql}

The cuts described above are implemented through the following ADQL query, which can be executed in the {\it Gaia} archive. 

\begin{lstlisting}
select * from gaiadr3.gaia_source
where phot_g_mean_mag < 20.5
and bp_rp < 0.5
and pm > 50
and ruwe < 1.4
and ( (4.74*pm/parallax > 600 and 4.74*pm/(parallax + parallax_error) > 400) or parallax < 0 )
\end{lstlisting}

The query is similar to the one employed by \citetalias{El-Badry2023}, with the main difference being that the magnitude limit is $G = 20.5$ here and was $G = 20.0$ in that work. We also remove the cut of \texttt{parallax\_over\_error} $<5$ adopted by \citetalias{El-Badry2023}, which results in the addition of a few relatively bright candidates not considered in that work.  

The query returns 92 sources, of which 25 were already considered by \citetalias{El-Badry2023}. We provide a complete log of the candidates and our follow-up in Appendix~\ref{sec:appendix}. We have characterized all 92 candidates via a combination of our own spectroscopic follow-up and analysis of archival data. Because the classifications and radial velocities (RVs) require spectroscopic information, we summarize the outcome of the survey after describing the follow-up data in Section~\ref{sec:summary_follow_up}. 

\section{Follow-up}
\label{sec:spectra}

We obtained dedicated follow-up spectroscopy of all candidates, except in cases where archival data could rule out a source as a viable supernova survivor.  All spectra from this follow-up program are published together with this paper.\footnote{\href{https://doi.org/10.22002/aff3x-28035}{https://doi.org/10.22002/aff3x-28035}} We describe our observations and analysis of survey data below. 

\subsection{Instruments}

\subsubsection{MagE}
\label{sec:MagE}
We observed 4 sources using the Magellan Echellette spectrograph \citep[MagE;][]{Marshall2008}  on the 6.5\,m Magellan Baade telescope at Las Campanas Observatory. All observations were carried out with the $0.7''$-wide slit, yielding a spectral resolution $R \approx 5500$ and wavelength coverage of 3500--11,000\,\AA. 

We reduced the spectra using \texttt{PypeIt} \citep[][]{Prochaska_2020}, which performs bias and flat-field correction, cosmic-ray removal, wavelength calibration, sky subtraction, extraction of 1D spectra, merging of spectral orders, and heliocentric RV corrections. 

\subsubsection{LRIS}
We observed 25 sources using the Low Resolution Imaging Spectrometer \citep[LRIS;][]{Oke1995} on the 10\,m Keck-I telescope on Maunakea. Some of the observations used the 600/7500 grating on the red side and the 600/4000 grism on the blue side, resulting in a full width at half-maximum intensity (FWHM) of 4.7\,\AA\ on the red side and 4.0\,\AA\ on the blue side, or a typical resolution $R\approx 1500$. Other observations used the 400/8500 grating on the red side and the 400/3400 grism on the blue side, resulting in FWHM $\approx 6.9$\,\AA\ on the red side and 6.5\,\AA\ on the blue side, or a typical resolution  $R\approx 1000$. LRIS is equipped with an atmospheric dispersion corrector.

We reduced the data using the LRIS automated reduction pipeline \citep[LPIPE;][]{Perley2019}, which performs bias and flat-field corrections, cosmic-ray removal, wavelength calibration and flexure corrections using night-sky lines, extraction of 1D spectra, telluric corrections, and flux calibration using a standard star. 

\subsubsection{GMOS}
\label{sec:GMOS}
We observed 6 sources using the Gemini Multi-Object Spectrograph \citep[GMOS;][]{Hook2004} on the 8.1\,m Gemini-South telescope at Cerro Pachón (programs GS-2023A-FT-107, GS-2023B-Q-139, and GS-2026A-FT-204). We used the B1200\_G5321 grating with a $0.75''$-wide slit and central wavelength of 4500\,\AA, leading to wavelength coverage from 3700 to 5300\,\AA\ and resolution $R\approx 3000$. We obtained a spectrum of a CuAr comparison lamp for wavelength calibration on-sky immediately after each 900\,s science exposure. The data were reduced using \texttt{PypeIt}.

\subsubsection{SOAR}
We observed 32 sources with the Goodman Spectrograph \citep{Clemens2004} on the SOAR telescope.
All SOAR observations used a 1.2\arcsec\ slit and a 400\,line\,mm$^{-1}$ grating, giving an approximate wavelength coverage of 4000--7850\,\AA\ with a FWHM resolution of $\sim 6.6$\,\AA. For each candidate we obtained a single exposure of length 1200--1800\,s, using longer exposures for fainter targets. All spectra were reduced and optimally extracted in the standard manner. The wavelength calibration was performed with comparison-lamp spectra obtained immediately after each science exposure. 

\subsubsection{DBSP}
We observed 2 sources with the Double Spectrograph \citep[DBSP;][]{Oke1982} on the 5\,m Hale telescope at Palomar Observatory. The 600/4000 grating was used on the blue side and the 316/7500 grating on the red side, together with the D55 dichroic. We employed the $1.0''$ or $1.5''$ slit, depending on the seeing, yielding an average resolution $R\approx 1500$ and wavelength coverage of 3000--10,700\,\AA. 

We reduced the spectra using \texttt{PypeIt} \citep[][]{Prochaska_2020} with the DBSP\_DRP wrapper \citep{Mandigo-Stoba2022}. This performs bias and flat-field corrections, cosmic-ray removal, wavelength calibration, extraction of 1D spectra, telluric corrections, and flux calibration using a standard star. 

\subsubsection{X-shooter}
\label{sec:XSHOOTER}

We observed 5 sources using the X-shooter spectrograph \citep[][]{Vernet2011} on the 8.2\,m UT3 telescope at the VLT on Cerro Paranal (program 115.27UU.001). We used the 0.8, 0.9, and 0.6 arcsec slits on the UVB, VIS, and NIR arms. This setup yielded a resolution of $R\approx 9,700$ in the UVB arm, $R\approx 18,400$ in the VIS arm, and $R\approx 11,600$ in the NIR arm, with near-continuous wavelength coverage from 3100--24000\,\AA. We reduced the data using the ESO Reflex pipeline \citep{Freudling2013} with standard calibrations.

\subsubsection{ESI}
\label{sec:ESI}
We observed 3 sources with the Echellette Spectrograph and Imager \citep[ESI;][]{Sheinis2002} on the 10\,m Keck-II telescope on Maunakea. We used the $0.5''$-wide slit, yielding a resolution $R\approx 8000$, with a useful wavelength coverage of 3900--10,000\,\AA. We reduced the data using the Mauna Kea Echelle Extraction (\texttt{MAKEE}) pipeline, which performs bias subtraction, flat fielding, wavelength calibration, and sky subtraction. The pipeline also carries out a linear shift to the wavelength solution using night-sky emission lines. 

\subsubsection{Gaia XP}

We identified 7 bright candidates ($G\lesssim 16$) as A and F-type stars based on the presence of strong Balmer lines in their {\it Gaia} XP spectra (\citealt{DeAngeli2023}; we visualize these using the \texttt{GaiaXPy} tool with \texttt{truncation = true}). The sources have well-measured parallaxes and fall on the main sequence of the CMD. These sources could be hypervelocity stars launched by the \citet{Hills1988} mechanism \citep[e.g.][]{Brown2015, Koposov2020}, ejection from binaries in core-collapse supernovae \citep[e.g.][]{Blaauw1961, Renzo2019}, or dynamical interactions in dense clusters \citep[e.g.][]{Leonard1991}, or they could be ordinary halo stars in the tail of the $v_\perp$ distribution for objects bound to the Milky Way. In either case, the presence of strong hydrogen lines in their spectra makes them unlikely to be survivors of double-degenerate supernovae. 

\subsubsection{Other surveys}
A few sources were also observed by SDSS spectroscopic surveys \citep[e.g.][]{Yanny2009, Kleinman2013}, by the LAMOST survey \citep[][]{Cui2012}, and by other works targeting halo stars \citep[e.g.][]{Brown2008}.

\subsection{Classification and RV measurements}
We visually inspected all spectra, comparing them to templates for normal stars, WDs of several spectral types, and known \Dsix and LP 40-365 stars. Results of this classification are listed in Table~\ref{tab:query}.

We measured RVs from pseudo-continuum normalized spectra by minimizing $\chi^2$ between the data and Doppler-shifted template spectra. For main-sequence contaminants, we used synthetic spectra from the BOSZ grid \citep{Bohlin2017} calculated with ATLAS and SYNTHE \citep{Kurucz1993}. For DA WD contaminants, we used  templates from \citet{Koester2010}; for DB WDs, we used templates from \citet{Cukanovaite2021}. For \Dsix and LP 40-365 stars, we used the C/O-dominated atmosphere library described by \citetalias{El-Badry2023}; these are tailored to hydrogen-free, metal-rich \Dsix atmospheres calculated with ATLAS and SYNTHE.  

We applied barycentric corrections using the mid-exposure times and target coordinates. For SOAR and DBSP spectra, we correct for flexure by measuring a residual wavelength zero-point offset using telluric absorption lines centered on the atmospheric A band  \citep[e.g.][]{Griffin1973, Nagarajan2023}. We adopted minimum RV uncertainties of $50\,{\rm km\,s^{-1}}$ for the SOAR, DBSP, and LRIS spectra. For most objects, this exceeds the formal RV uncertainty and is meant to account for possible shifts in the wavelength solution due to flexure. Most of our confirmed hypervelocity runaways were followed-up with more stable, higher-resolution spectrographs, allowing us to measure more precise RVs.

\subsection{Contaminants}
We summarize the several classes of false positives we rejected below. 

\subsubsection{Normal WDs and hot subdwarfs}
The most common contaminants are WDs and hot subdwarfs (``sdO/B'' stars). While some of these could conceivably also be thermonuclear SN runaways, they have comparatively low RVs, and are most likely bound halo objects with underestimated parallaxes and overestimated $v_\perp$. We label these objects according to their spectral type in Table~\ref{tab:query}.

\subsubsection{Main-sequence K dwarfs}
Spectroscopic follow-up showed some candidates to be main-sequence K dwarfs, usually low-metallicity halo stars. While such sources normally should fall redward of our color cut, noisy {\it Gaia} colors for faint sources cause some redder sources to scatter into the sample. Such sources are labeled ``MS'' in Table~\ref{tab:query}. 

\subsubsection{Main-sequence A and F stars}

About a dozen candidates are relatively bright ($G=10-16$) and have well-constrained parallaxes. Most of these sources have spectra from archival surveys and did not require additional follow-up. Their spectra show Balmer lines characteristic of A and F-type stars, consistent with their position on the main sequence of the {\it Gaia} CMD. The sources' CMD positions and hydrogen-dominated spectra make it unlikely that these sources are runaways from WD+WD binaries, though some may still be high-velocity runaways launched through other channels. Such sources are also labeled ``MS'' in Table~\ref{tab:query}.

\subsubsection{Blended sources}
Several candidates are faint sources within a few arcseconds of a much brighter source. Since such sources are overrepresented in our candidate sample relative to the full {\it Gaia} catalog and there is no astrophysical reason to expect thermonuclear runaways to be preferentially found near a bright companion, we suspect that they have spurious colors and/or astrometry due to blending \citep[e.g.][]{El-Badry2021_wbs, Rybizki2022}. Such sources are marked ``blended'' in Table~\ref{tab:query}; in most cases, we did not follow them up spectroscopically. 

\subsubsection{Wide binaries}
A few candidates have a comparably bright companion within a few arcsec with near-identical high proper motion, making them very likely halo wide binaries. In some cases, the colors and astrometry are likely affected by blending; in others, the companion is distant enough that blending is unlikely to be very important. The high accelerations involved in launching hypervelocity runaways would unbind any wide binaries. These sources are marked ``WB'' in Table~\ref{tab:query} and were not followed-up.

\subsection{Results}
\label{sec:summary_follow_up}

\begin{figure*}
    \centering
    \includegraphics[width=\textwidth]{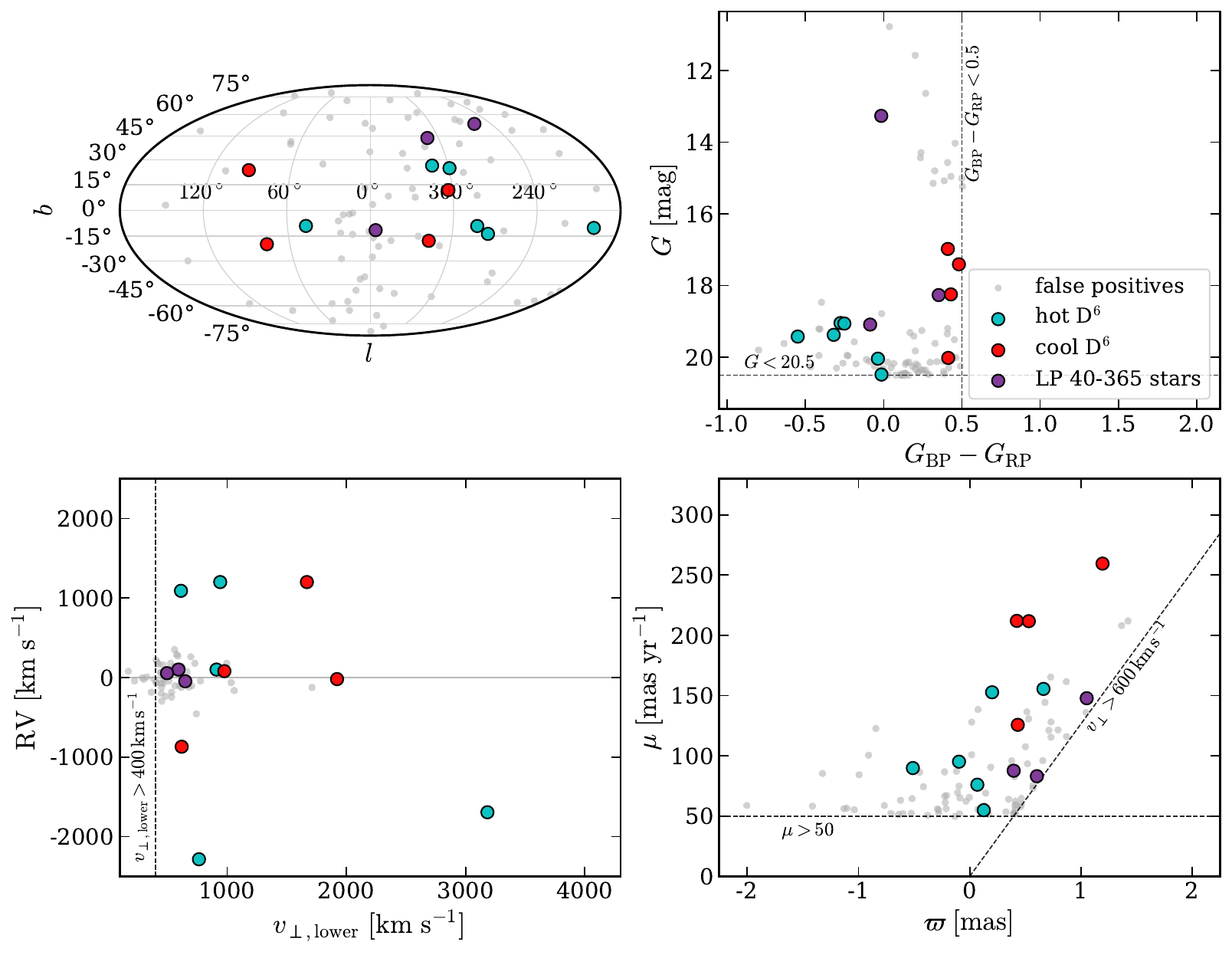}
    \caption{Basic properties of the 92 candidates returned by our {\it Gaia} search. Gray points show false positives, while colored points show different classes of spectroscopically confirmed runaways: hot \Dsix stars (cyan), cool \Dsix stars (red), and LP~40-365 stars (purple). Panels show sky positions, the {\it Gaia} color--apparent magnitude diagram, radial velocity as a function of the $1\sigma$ tangential-velocity lower limit $v_{\perp,{\rm lower}}=4.74\mu/(\varpi+\sigma_\varpi)$, and the parallax--proper-motion plane. The RV panel includes only sources with measured RVs. Dashed black lines show our selection cuts on color, magnitude, proper motion, and tangential velocity. Relative to the initial candidates, the confirmed runaways are found at lower Galactic latitudes, brighter apparent magnitudes, and higher RV amplitudes. The LP 40-365 stars are found at lower velocities than the \Dsix stars.}
    \label{fig:search_sample}
\end{figure*}

\begin{table}
\centering
\caption{Candidate accounting for the {\it Gaia} search.}
\label{tab:survey_accounting}
\begin{tabular}{lr}
\hline
Category & Number \\
\hline
Sources returned by ADQL query & 92 \\
Previously studied by \citetalias{El-Badry2023} & 25 \\
\hline
Spectroscopically confirmed runaways & 13 \\
\quad \Dsix stars & 10 \\
\quad LP~40-365 stars & 3 \\
\hline
False positives & 79 \\
\quad WDs and hot subdwarfs & 45 \\
\quad Main-sequence stars & 20 \\
\quad Blends and wide binaries & 14 \\
\hline
\end{tabular}
\tablecomments{Confirmed runaways are classified based on follow-up spectroscopy. False positives are candidates rejected through our follow-up, archival spectra, or inspection of the {\it Gaia} astrometry. A detailed log is provided in Table~\ref{tab:query}.}
\end{table}

Table~\ref{tab:survey_accounting} summarizes our classifications. The sample of 92 candidates yields 13 confirmed thermonuclear SN runaways and 79 false positives. Seven of the confirmed runaways were already recovered by \citetalias{El-Badry2023}, including six objects discovered in that work and one previously discovered by \citet{Shen2018}. Three of the remaining six are new discoveries, while the other three were discovered by \citet{Shen2018} and \citet{Raddi2019} but not recovered by \citetalias{El-Badry2023}.

Figure~\ref{fig:search_sample} shows basic properties of sources in both groups. Both false positives and confirmed runaways are found primarily at high latitude relative to the bulk of the {\it Gaia} catalog; this is mainly due to our requirement of $G_{\rm BP}-G_{\rm RP} < 0.5$, which sets an effective maximum extinction of $A_V \lesssim 2$ mag. Confirmed runaways are still found at lower latitude on average than false positives, reflecting the fact that they are launched primarily from the Galactic disk and remain bright for at most a few Myr (Section~\ref{sec:space_density}). In contrast, most of the false positives are halo stars, which are distributed more isotropically.

The false positives separate into two broad groups in color-magnitude space. The bright group is dominated by A- and F-type main-sequence stars with well-measured positive parallaxes and tangential velocities just above the $600\,{\rm km\,s^{-1}}$ threshold. These objects are likely metal-poor halo stars or main-sequence binary runaways. The faint group is dominated by WDs and faint halo stars with large parallax uncertainties. Noisy parallaxes can scatter ordinary halo objects in this group into the high-$v_\perp$ selection, and such contaminants become increasingly prevalent at faint magnitudes. 

Confirmed runaways have systematically larger absolute RVs than false positives, though some still have RVs close to zero. All the LP 40-365 stars detected by our search have $\left | \rm RV \right | < 100\, {\rm km\,s^{-1}}$ ; this reflects the fact that LP 40-365 stars are ejected with typical velocities of $\sim 600\,{\rm km\,s^{-1}}$, so any sources satisfying our cut of $v_\perp > 600\,{\rm km\,s^{-1}}$ (Section~\ref{sec:space_density}) must necessarily have low RVs. We can approximately estimate the fraction of sources missed due to our tangential velocity cut as follows. For randomly oriented trajectories, the ratio $x=v_\perp/v_{\rm 3D}$ is distributed as $p(x)=x/\sqrt{1-x^2}$ \citep{Nottale2018}. This implies that populations with true velocities of 650, 1000, and $2000\,{\rm km\,s^{-1}}$ would have approximately 62\%, 20\%, and 5\% of their members with $v_{\perp} < 600\,{\rm km\,s^{-1}}$, respectively. This simplified estimates neglects effects of Galactic rotation and our additional cut on $v_{\perp,\,\rm lower}$, but the basic conclusion is that projection effects will cause our search to miss only a small fraction of high-velocity \Dsix stars, but an order-unity fraction of lower-velocity LP~40-365 stars. We refer to \citet{Raddi2019} for further discussion of this issue.

A few candidates -- all of which turned out to be false positives -- have $v_{\perp,{\rm lower}}<400\,{\rm km\,s^{-1}}$ in the lower-left panel; these are sources with negative measured parallaxes, which satisfy the final ``or \texttt{parallax < 0}'' branch of the ADQL query.  

\section{New discoveries}
\label{sec:discoveries}

Our search yielded three new \Dsix stars, whose properties we discuss below. 

\subsection{J1251-5059}
\label{sec:j1251}

J1251-5059 ({\it Gaia} DR3 6078434976557457536) is a cool \Dsix star with a spectrum very similar to the three cool  \Dsix stars discovered by \citet{Shen2018}. We first identified the source as a likely \Dsix star based on a SOAR spectrum and subsequently obtained higher-resolution spectra with GMOS and X-shooter. 

The source's zeropoint-corrected {\it Gaia} DR3 parallax, $\varpi = 0.45 \pm 0.53$, is not particularly constraining, but it rules out distances $d < 1.0$\,kpc ($d < 0.7$\,kpc) at the $1\sigma$ ($2\sigma$) level.  The 3D dust map of \citet{Wang2025} predicts $E(B-V) = 0.15\pm 0.02$ across this distance range, while the \citet{Schlegel1998} map predicts $E(B-V) = 0.18$ at infinity. The best-fit intrinsic color of $(G_{\rm BP} - G_{\rm RP})_0 = 0.20 \pm 0.10$ mag is slightly bluer than the three \Dsix stars from \citet{Shen2018}, but the source's color is consistent with the colors of all three systems within $2\sigma$. The color uncertainty is dominated by photon noise.

The three cool \Dsix stars from \citet{Shen2018} -- which have better-constrained parallaxes than J1251-5059 -- have intrinsic $G$-band absolute magnitudes ranging from 7.2 to 5.8. Requiring J1251-5059 to fall within this absolute magnitude range would require a distance between 3.0 and 5.7 kpc, while any distance below 3.0 kpc would make J1251-5059 the least luminous cool \Dsix star discovered so far. 

Comparing the X-shooter spectrum to Kurucz models, we measure an RV of $-870\pm 5\,{\rm km\,s^{-1}}$. The source's proper motion implies a tangential velocity $v_{\perp}=1788\,{\rm km\,s^{-1}}\times\left[d/\left(3\,{\rm kpc}\right)\right]$. Although the source's total velocity is uncertain, it thus seems likely that it has a total velocity of $\gtrsim 1500\,{\rm km\,s^{-1}}$. Our kinematic modeling (Section~\ref{sec:orbits}) leads to an inferred ejection velocity of $1508^{+838}_{-413}\,{\rm km\,s^{-1}}$.

\begin{figure*}
    \centering
    \includegraphics[width=\textwidth]{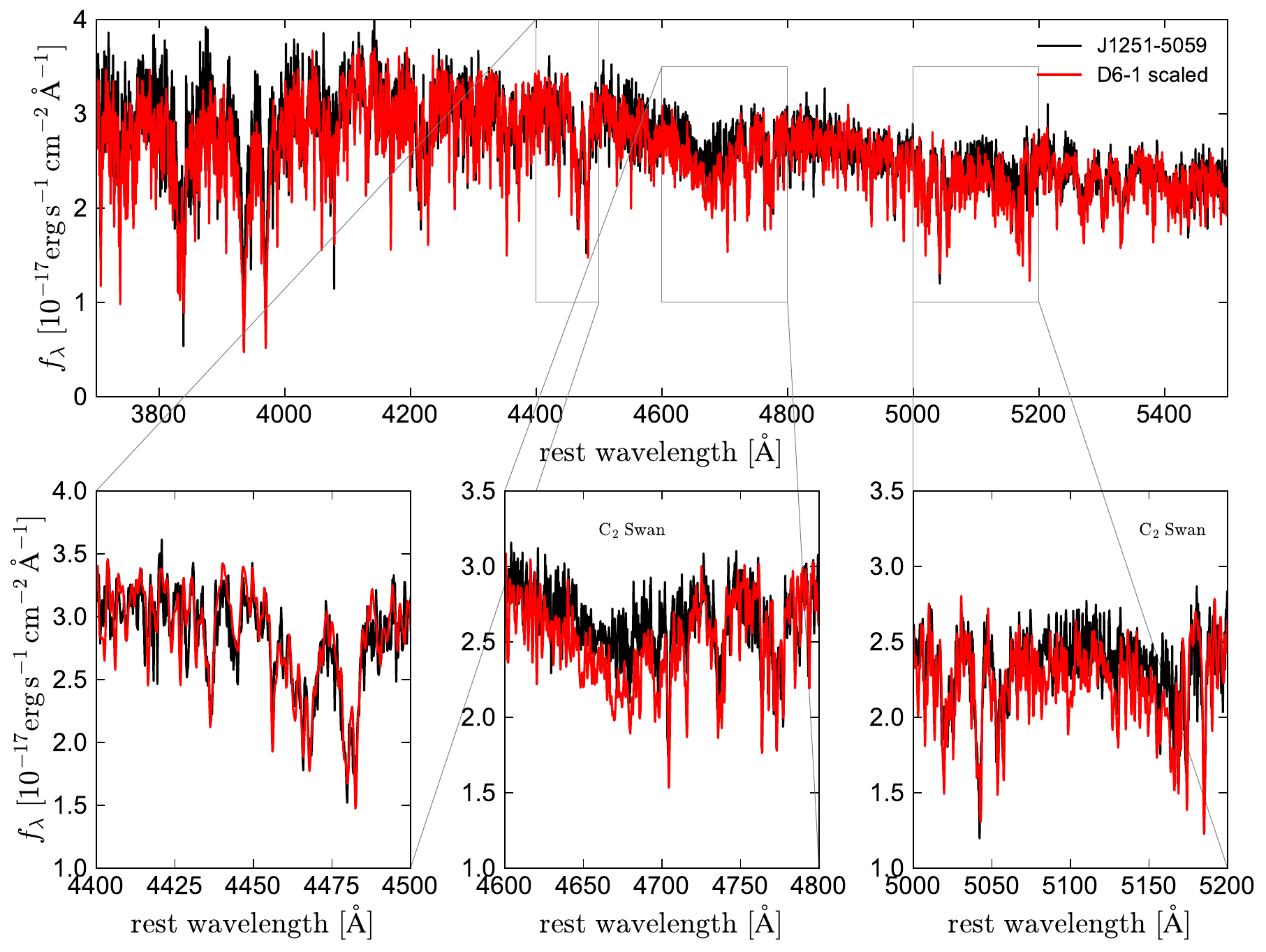}
    \caption{Comparison of the X-shooter spectra of J1251-5059, a new \Dsix star discovered by our search (black) and D6-1, a brighter and nearer source discovered by \citet{Shen2018}. The spectrum of D6-1 is multiplied by 0.1 so that the flux scales of the two sources are consistent. Overall, the two stars have extremely similar spectra after this rescaling, indicating that they have similar temperatures and surface compositions. However, there are two broad absorption features that are significantly deeper in D6-1 than in J1251-5059. These are likely Swan bands due to molecular carbon, implying that  J1251-5059 has a lower carbon abundance than D6-1.  }
    \label{fig:J1251_spectrum}
\end{figure*}

In  Figure~\ref{fig:J1251_spectrum}, we compare the X-shooter spectrum of J1251-5059 to an X-shooter spectrum of the star D6-1. That spectrum was obtained on July 11, 2018 (program 0101.C-0646), and will be subject to a detailed abundance analysis in future work. Both spectra are shifted to rest frame. 
We have rescaled the spectrum of D6-1 by a factor of 0.1 so that the fluxes of the two sources are on the same scale; this corresponds to moving D6-1 to a 3.16 times larger distance at fixed radius. 

The spectrum of J1251-5059 is very similar to that of D6-1. Both stars have spectra dominated by metal lines, including Mg, Si, Ca, O, and Na. The objects' similar spectra suggest that they have both similar atmospheric parameters and compositions. The most important difference between the two objects' spectra is in the strength of two broad absorption features near 4700 and 5150\,\AA. These are very likely the Swan bands due to molecular carbon (C$_2$), which are often observed in the spectra of cool WDs with carbon in their atmospheres \citep[e.g.][]{Schmidt1995}. As discussed by \citet{Werner2024} and \citet{Hollands2025}, the ratio of carbon to oxygen in the atmospheres of known \Dsix stars varies by an order of magnitude. Systems with higher carbon abundances have stronger C$_2$ bands, implying that J1251-5059 contains less carbon relative to oxygen than D6-1. 

J1251-5059 was flagged as a variable star by the {\it Gaia} DR3 variability processing pipeline \citep{Eyer2023}. We retrieved the source's $G-$band light curve from the DR3 DataLink server. It contains 49 usable flux measurements taken over a period of 998 d. The source appears variable at the 0.1 mag level, with scatter of about twice the amplitude of the typical per-epoch uncertainties. This variability is reminiscent of the rotational modulation observed in D6-2 \citep{Chandra2022}. We are unable to unambiguously identify a variability period from the light curve, but it may be possible to identify a rotation period for the source with a light curve from {\it Gaia} DR4 or Rubin/LSST.

\subsection{J0812-5943}
\label{sec:j0812}

J0812-5943 ({\it Gaia} DR3 5290552672404889984) is a blue \Dsix star with a spectrum more similar to the hot objects discovered by \citetalias{El-Badry2023} than to the cool objects from \citet{Shen2018}. We identified the source as a promising candidate from a SOAR spectrum and subsequently obtained several higher-SNR spectra with GMOS.

The source's zeropoint-corrected {\it Gaia} DR3 parallax, $\varpi = -0.40 \pm 0.57$, favors large distances; $d < 1.6\,{\rm kpc}$ is ruled out at the $2\sigma$ level.  The 3D dust map of \citet{Wang2025} predicts $E(B-V) = 0.17\pm 0.02$ across $1.5-5$\,kpc, while the \citet{Schlegel1998} map predicts $E(B-V) = 0.20$ at infinity. These predictions lead to a predicted intrinsic color of $(G_{\rm BP}-G_{\rm RP})_0 \approx -0.25\pm 0.23$, with the uncertainty dominated by shot noise in the BP and RP fluxes. 

Among the currently known \Dsix stars, the spectrum of J0812-5943 is most similar to J0927-6335, a source with an inferred absolute magnitude of $6 \lesssim M_{G,0} \lesssim 7$ \citepalias{El-Badry2023}. J0812-5943 would fall in this absolute magnitude range for distances of $4-6$ kpc. At these distances, the source's proper motion implies a tangential velocity $v_{\perp}=1704\,{\rm km\,s^{-1}}\times\left[d/\left(4\,{\rm kpc}\right)\right]$. We measure a radial velocity of $-1210\pm 10\,{\rm km\,s^{-1}}$. These velocities are again suggestive of a high space velocity of order $2000\,{\rm km\,s^{-1}}$ but are subject to significant uncertainties due to the source's poorly constrained parallax. Our kinematic modeling (Section~\ref{sec:orbits}) leads to an inferred ejection velocity of $1924^{+674}_{-459}\,{\rm km\,s^{-1}}$.

To obtain rough constraints on the source's temperature and surface gravity, we compared the spectrum of J0812-5943 to the C/O-dominated atmosphere models calculated by \citetalias{El-Badry2023} for hot \Dsix stars. Those models were computed with the Kurucz ATLAS/SYNTHE machinery, using hydrogen-free, C/O-rich compositions with enhanced metal abundances motivated by the observed spectra of \Dsix stars. \citetalias{El-Badry2023} calculated a library of models spanning effective temperatures $15,000 < T_{\rm eff}/{\rm K} < 150,000$ and surface gravities $6 < \log\left[g/\left({\rm cm\,s^{-1}}\right)\right] < 8$; J0812-5943 is best-matched by the model with $T_{\rm eff} = 49,000$\,K and $\log\left[g/\left({\rm cm\,s^{-1}}\right)\right] = 7.4$.

\begin{figure*}
    \centering
    \includegraphics[width=\textwidth]{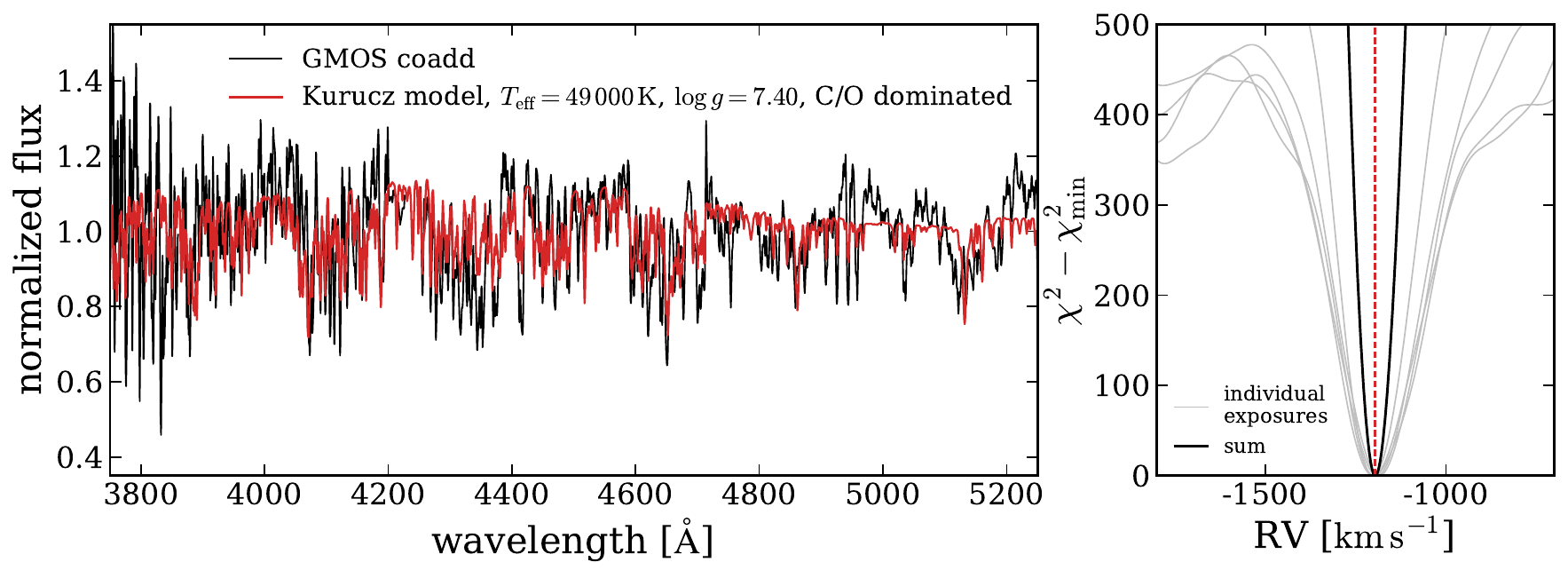}
    \caption{GMOS spectrum and RV constraints for J0812-5943, a new hot \Dsix star discovered by our search. Left: coadded, normalized GMOS spectrum (black) compared to a normalized Kurucz model with $T_{\rm eff}=49{,}000\,{\rm K}$, $\log g=7.40$, and a C/O-dominated composition (red). Right: $\chi^2-\chi^2_{\rm min}$ as a function of radial velocity. Gray curves show the individual exposures, while the black curve shows their sum. The best-fit radial velocity is $\rm RV = -1210\pm 10\,km\,s^{-1}$.}
    \label{fig:J0812_spectrum}
\end{figure*}

Figure~\ref{fig:J0812_spectrum} compares the coadded GMOS spectrum to this model. The right panel shows the RV constraint from the individual exposures and from a joint analysis of all exposures. The model contains most of the strong lines present in the observed spectrum, allowing a robust measurement of the star's RV and supporting our classification of it as a hot \Dsix star. The star is very likely cooler than J0927-6335 -- the source's closest spectral cousin among the currently known \Dsix stars --  because its O III and C III lines are significantly stronger relative to O IV and C IV. For J0927-6335, \citet{Werner2024} found $T_{\rm eff} = 60,000\,{\rm K}$.

We adopt $T_{\rm eff} = 49,000$\,{\rm K} as our best estimate of the star's temperature, but this should be interpreted with caution until a full analysis with bespoke atmospheric models is possible. Our Kurucz atmosphere models are suitable for identifying the dominant species and measuring RVs, but they are LTE models computed with a code that is not well tested for sources with atmospheric parameters and compositions like the hot \Dsix stars. It is clear from Figure~\ref{fig:J0812_spectrum} that the model is an imperfect fit to the data. This was also true for the objects studied by \citetalias{El-Badry2023} and likely reflects both limitations of the model spectra and a mismatch between the true and assumed abundance patterns. We also experimented with fitting the spectrum with NLTE models produced with the TMAP code (see Section~\ref{sec:j1949}), but we found that the Kurucz code produces a significantly better fit for this star, likely because the line list used for the TMAP analysis lacks many of the O III and O IV lines present in the Kurucz model. 

\subsection{J1949+0745}
\label{sec:j1949}
J1949+0745 ({\it Gaia} DR3 4298029440184716160) is another hot \Dsix star. The source's zeropoint-corrected {\it Gaia} DR3 parallax, $\varpi = 0.20 \pm 0.59$, favors large distances, ruling out $d < 1.3$\,kpc ($d < 0.7$\,kpc) at the $1\sigma$ ($2\sigma$) level. The 3D dust map of \citet{Wang2025} predicts $E(B-V) = 0.25\pm 0.01$ across $1.5-5$\,kpc, while the \citet{Schlegel1998} map predicts $E(B-V) = 0.27$ at infinity. These predictions lead to a predicted intrinsic color of $(G_{\rm BP}-G_{\rm RP})_0 \approx -0.37\pm 0.12$, consistent with the Rayleigh-Jeans tail of a hot star. The source's proper motion implies a tangential velocity $v_{\perp}=2172\,{\rm km\,s^{-1}}\times\left[d/\left(3\,{\rm kpc}\right)\right]$. The RV is relatively low, with a best-fit value of $\rm RV = 100\pm 20\,{\rm km\,s^{-1}}$.

\begin{figure*}[!t]
    \centering
    \includegraphics[width=\textwidth,trim=0 0 0 3.1in,clip]{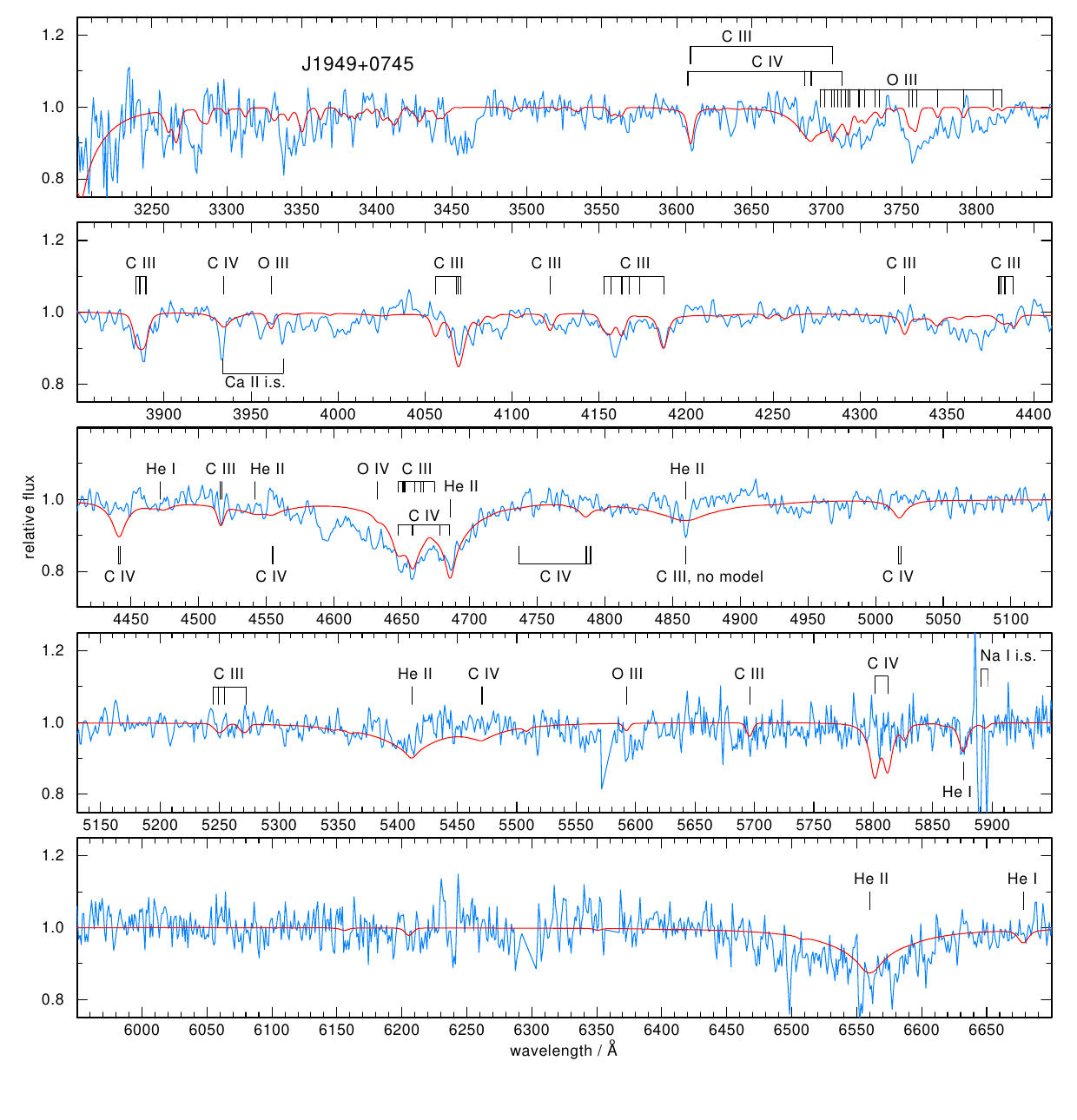}
    \caption{Normalized spectrum of J1949+0745 compared to a He-dominated NLTE atmosphere model calculated with TMAP.  The model has $T_{\rm eff}=65{,}000\,{\rm K}$, $\log g=8.5$, and abundances He = 0.58, C = 0.25, and O = 0.17 (mass fractions). It reproduces most of the strongest lines, supporting classification of J1949+0745 as a hot \Dsix star, but the model does not fully match all features in the spectrum. In particular, the C~IV lines near 4440\,\AA\ and 5802/5812\,\AA\ are stronger in the model than in the data.}
    \label{fig:J1949_spectrum}
\end{figure*}

Figure~\ref{fig:J1949_spectrum} shows the LRIS spectrum of J1949+0745, coadded across three exposures. The spectrum resembles that of a PG~1159 star, being dominated by lines of C~IV and He~II \citep[e.g.][]{Werner2006}. Among the known sample of \Dsix stars, it is most similar to J0927-6335. It is likely hotter than J0812-5943, with stronger C~IV relative to C~III. Unlike J0927-6335 and J0812-5943, it also shows He lines. 

Figure~\ref{fig:J1949_spectrum} compares the normalized spectrum to a NLTE atmosphere model calculated with TMAP \citep{Werner2003}. The model is He-dominated (He = 0.58, C = 0.25, O = 0.17, mass fractions), with $T_{\rm eff}=65{,}000\,{\rm K}$, $\log g=8.5$. 

Cooler models produce He~I lines that are too strong (the strongest being He~I $\lambda5876$\,\AA), while hotter models lose the observed C~III features. Lower surface gravities overpredict the cores of the higher He~II series lines.

The strongest tension in this model is in the C~IV features near 4440\,\AA\ and 5802/5812\,\AA, which are stronger than observed. Reducing the carbon abundance could improve those lines but would underpredict the C-rich absorption trough near 4650--4660\,\AA. There is evidence for oxygen: The strongest feature predicted by the model is a O~III line blend around 3750\,\AA, but it is not strong enough to fit the observation. Models with higher O abundances or lower temperatures would fit this feature better. However, higher O abundances cause other O~III lines to become too strong (in particular $\lambda3962$\,\AA), while lower temperatures lead to a poorer fit to the He~I lines. The spectrum therefore suggests that J1949+0745 may be a He-dominated PG 1159 star, perhaps representing systems like the source J1332-3541 \citep{Werner2024} but with a different mass or evolutionary phase. Test models show that the hydrogen abundance in J1949+0745 must be lower than 10\,\%, otherwise the H$\beta$/He~II $\lambda4860$ line blend would be significantly too strong in the model. We consider it unlikely that J1949+0745 is an ordinary PG 1159 star with an underestimated parallax, because a normal WD with the star's parameters would have already crossed the wind limit and transitioned to become a DO WD \citep[e.g.][]{Unglaub2000, Bedard2022, Mackensen2025}.

\section{Kinematic modeling}
\label{sec:orbits}

We jointly infer distances, velocities, and kinematic ages of all objects discovered by our search, and for other \Dsix and LP 40-365 stars from the literature, following the procedure described by  \citetalias{El-Badry2023}. For each source, we sample the posterior over absolute magnitude, proper motion, and RV using the measured {\it Gaia} astrometry, RV, their uncertainties, and the {\it Gaia} astrometric covariance matrix; we also apply the {\it Gaia} parallax zeropoint correction. Our fiducial distance prior is flat in absolute magnitude, $M_G$. This is equivalent to a distance prior $p(d) \propto 1/d$, downweighting large distances and velocities. We also adopt a prior that the total ejection velocity cannot exceed $3000\,{\rm km\,s^{-1}}$. Sensitivity to the adopted distance prior is explored by \citetalias{El-Badry2023}. For US 708, we adopt a Gaussian distance prior of $d=8.5\pm 1\,{\rm kpc}$ following \citet{Geier2015}.

For each posterior sample, we transform the heliocentric phase-space coordinates to a Galactocentric frame and compute the current height above the Galactic plane and total velocity. We then integrate the orbit in the \texttt{MWPotential2014} Galactic potential using \texttt{galpy} \citep{Bovy2015}. This is an improvement over the treatment of \citetalias{El-Badry2023}, who neglected the gravitational potential and assumed straight-line orbits. 
We integrate backward or forward in time, depending on the sign of the initially inferred midplane-crossing time, and identify the most recent or next crossing of the disk midplane. We define the ejection velocity, $v_{\rm ejection}$, by subtracting the local circular velocity of the Galactic potential at this crossing. For most objects, this potential-integrated treatment changes the inferred ejection velocity and midplane-crossing time only modestly relative to the straight-line approximation used by \citetalias{El-Badry2023}, but the difference is non-negligible for the longest flight-time systems. Most notably, the inferred ejection velocity of US 708 increases from $743^{+19}_{-15}\,{\rm km\,s^{-1}}$ when the potential is neglected to $802^{+14}_{-13}\,{\rm km\,s^{-1}}$ when it is accounted for. 

\begin{table*}[!t]
\begin{tabular}{llllllllll}
Name & {\it Gaia} DR3 Source ID &  $G$  & $G_{\rm BP}-G_{\rm RP}$ & $\rm RV$  & 
$d$ & $v_{\rm tot}$ & $v_{\rm ejection}$ & $z$ & $t_{\rm midplane}$ \\
  &   &  [mag] &  [mag] & $\rm [km\,s^{-1}]$  & 
$[\rm kpc]$ & $[\rm\,km\,s^{-1}]$ & $[\rm\,km\,s^{-1}]$ & $[\rm kpc]$ & $[\rm Myr]$ \\
\hline
\hline 
\multicolumn{10}{l}{\underline{Suspected $\rm D^6$ stars}}   \\ 
D6-1 & 5805243926609660032 & 17.4 & 0.48 & $1200 \pm 40$ & $1.91^{+0.29}_{-0.22}$ & $2045^{+255}_{-190}$ & $2260^{+253}_{-189}$ & $-0.57^{+0.07}_{-0.09}$ & $0.63^{+0.04}_{-0.04}$ \\ 
D6-2 & 1798008584396457088 & 17.0 & 0.41 & $80 \pm 10$ & $0.84^{+0.05}_{-0.04}$ & $1152^{+59}_{-53}$ & $1059^{+62}_{-56}$ & $-0.27^{+0.02}_{-0.02}$ & $-0.72^{+0.01}_{-0.01}$ \\ 
D6-3 & 2156908318076164224 & 18.2 & 0.43 & $-20 \pm 80$ & $2.25^{+0.36}_{-0.36}$ & $2237^{+359}_{-354}$ & $2402^{+395}_{-390}$ & $0.93^{+0.14}_{-0.14}$ & $2.22^{+0.19}_{-0.16}$ \\ 
J1235 & 6156470924553703552 & 19.0 & -0.28 & $-1694 \pm 10$ & $4.07^{+1.01}_{-1.18}$ & $2672^{+340}_{-348}$ & $2471^{+353}_{-345}$ & $1.73^{+0.42}_{-0.49}$ & $1.87^{+0.83}_{-0.25}$ \\ 
J0927 & 5250394728194220800 & 19.4 & -0.32 & $-2285 \pm 20$ & $4.60^{+2.20}_{-1.77}$ & $2764^{+278}_{-157}$ & $2560^{+289}_{-168}$ & $-0.71^{+0.28}_{-0.35}$ & $-1.34^{+0.43}_{-0.42}$ \\ 
J0546 & 3335306915849417984 & 19.1 & -0.25 & $1200 \pm 20$ & $4.12^{+2.23}_{-1.77}$ & $1725^{+672}_{-414}$ & $1886^{+677}_{-435}$ & $-0.71^{+0.31}_{-0.39}$ & $0.61^{+0.05}_{-0.08}$ \\ 
J1332 & 6164642052589392512 & 19.4 & -0.55 & $1090 \pm 50$ & $1.64^{+1.23}_{-0.69}$ & $1474^{+764}_{-339}$ & $1632^{+735}_{-331}$ & $0.75^{+0.54}_{-0.31}$ & $0.64^{+0.14}_{-0.14}$ \\ 
{\bf J1251} & 6078434976557457536 & 20.0 & 0.41 & $-870 \pm 5$ & $2.16^{+1.67}_{-1.00}$ & $1563^{+826}_{-364}$ & $1508^{+838}_{-413}$ & $0.46^{+0.34}_{-0.20}$ & $-4.11^{+2.48}_{-11.32}$ \\ 
{\bf J0812} & 5290552672404889984 & 20.5 & -0.01 & $-1210 \pm 10$ & $3.68^{+1.91}_{-1.64}$ & $2144^{+659}_{-446}$ & $1924^{+674}_{-459}$ & $-0.86^{+0.39}_{-0.45}$ & $1.09^{+0.42}_{-0.12}$ \\ 
{\bf J1949} & 4298029440184716160 & 20.0 & -0.04 & $100 \pm 20$ & $2.15^{+1.26}_{-0.93}$ & $1733^{+912}_{-657}$ & $1529^{+913}_{-657}$ & $-0.32^{+0.15}_{-0.20}$ & $1.86^{+0.09}_{-0.16}$ \\ 
J1637 & 1327920737357113088 & 20.3 & -0.13 & $384 \pm 6$ & $3.10^{+3.17}_{-1.62}$ & $1188^{+940}_{-407}$ & $1058^{+1155}_{-445}$ & $2.09^{+2.11}_{-1.08}$ & $3.67^{+1.07}_{-1.21}$ \\

\multicolumn{10}{l}{\underline{Suspected LP 40-365 stars}}   \\ 
LP 40-365 & 1711956376295435520 & 15.6 & 0.23 & $498 \pm 5$ & $0.61^{+0.01}_{-0.01}$ & $837^{+6}_{-6}$ & $627^{+5}_{-5}$ & $0.43^{+0.01}_{-0.01}$ & $4.66^{+0.33}_{-0.29}$ \\ 
J1603 & 5822236741381879040 & 17.8 & 0.16 & $-480 \pm 5$ & $2.08^{+0.48}_{-0.33}$ & $835^{+62}_{-39}$ & $605^{+71}_{-42}$ & $-0.35^{+0.06}_{-0.09}$ & $1.56^{+0.12}_{-0.10}$ \\ 
J0905 & 688380457508503040 & 19.6 & 0.24 & $300 \pm 50$ & $4.60^{+7.43}_{-2.84}$ & $536^{+1099}_{-285}$ & $660^{+1068}_{-227}$ & $2.97^{+4.76}_{-1.82}$ & $-13.63^{+29.14}_{-12.78}$ \\ 
J1825 & 6727110900983876096 & 13.3 & -0.02 & $-47 \pm 5$ & $0.95^{+0.03}_{-0.02}$ & $429^{+19}_{-17}$ & $668^{+19}_{-17}$ & $-0.17^{+0.01}_{-0.01}$ & $1.59^{+0.02}_{-0.02}$ \\ 
J1311 & 3507697866498687232 & 18.3 & 0.35 & $55 \pm 10$ & $1.87^{+1.07}_{-0.50}$ & $949^{+422}_{-195}$ & $809^{+454}_{-211}$ & $1.31^{+0.74}_{-0.35}$ & $3.83^{+0.20}_{-0.20}$ \\ 
J1109 & 3804182280735442560 & 19.1 & -0.09 & $100 \pm 10$ & $2.84^{+2.30}_{-1.19}$ & $1357^{+949}_{-493}$ & $1155^{+983}_{-489}$ & $2.31^{+1.84}_{-0.96}$ & $3.88^{+0.28}_{-0.37}$ \\ 
J1022 & 5446737753669901568 & 19.0 & -0.22 & $-390 \pm 20$ & $7.52^{+7.25}_{-4.00}$ & $1300^{+945}_{-450}$ & $1208^{+997}_{-519}$ & $2.50^{+2.39}_{-1.32}$ & $2.79^{+0.53}_{-0.19}$ \\ 
J1240 & 1682129610835350400 & 18.4 & -0.29 & $-177 \pm 10$ & $0.42^{+0.02}_{-0.02}$ & $240^{+21}_{-19}$ & $395^{+17}_{-16}$ & $0.35^{+0.02}_{-0.02}$ & $-13.30^{+3.57}_{-5.36}$ \\ 

\multicolumn{10}{l}{\underline{Suspected runaway helium star donors}}   \\ 
US 708 & 815106177700219392 & 18.9 & -0.44 & $835 \pm 10$ & $8.38^{+1.02}_{-0.99}$ & $920^{+13}_{-12}$ & $802^{+14}_{-13}$ & $6.17^{+0.75}_{-0.73}$ & $12.51^{+2.00}_{-1.85}$ \\ 
\end{tabular}
\caption{Known high-velocity objects suspected to be runaways from thermonuclear supernovae. Uncertainties are $1\sigma$ (middle 68\%). Distances and velocities are inferred with a flat-$M_G$ prior. We distinguish between $v_{\rm tot}$, the current 3D space velocity in a Galactocentric frame, and $v_{\rm ejection}$, the assumed velocity with which the star was launched, assuming it was on a circular orbit in the Galactic disk at the time. $t_{\rm midplane} = z/v_z$ represents the time since the most recent midplane crossing.  Objects in bold are new discoveries. We also list previously known $\rm D^6$ stars \citep{Shen2018} and LP~40-365 stars \citep{Raddi2019}, as well as the one known high-velocity helium star US~708 \citep{Geier2015}. }
\label{tab:all_systems}
\end{table*}

Table~\ref{tab:all_systems} reports the results of applying this modeling to all known high-velocity runaways likely associated with thermonuclear supernovae. We discuss the population further in Section~\ref{sec:all_known}.

\subsection{Newly discovered objects}

J0812-5943 has an inferred distance of $d=3.68^{+1.91}_{-1.64}$ kpc. Accounting for extinction, this corresponds to $M_{G,0}\approx 7.2$, with an uncertainty of about 1 mag dominated by the distance uncertainty. This places it at the faint end of the hot \Dsix population, likely with a lower luminosity than J0927-6335 and J0546+0836. Its inferred 3D velocity is $v_{\rm tot}=2144^{+659}_{-446}\,{\rm km\,s^{-1}}$, and its inferred ejection velocity is $v_{\rm ejection}=1924^{+674}_{-459}\,{\rm km\,s^{-1}}$, reflecting a likely boost of $\sim 220\,{\rm km\,s^{-1}}$ due to Galactic rotation. J0812-5943 is below the midplane ($z=-0.86_{-0.45}^{+0.39}$ kpc) and moving away from it. It has a well-constrained midplane crossing time of $t_{\rm midplane}=1.09_{-0.12}^{+0.42}$ Myr.

J1949+0745 has a fiducial distance of $d=2.15^{+1.26}_{-0.93}$ kpc, implying $M_{G,0}\approx 7.7$. It is therefore similarly luminous to J0812-5943 and J1332-3541. We infer a total velocity of $v_{\rm tot}=1733^{+912}_{-657}\,{\rm km\,s^{-1}}$ and ejection velocity $v_{\rm ejection}=1529^{+913}_{-657}\,{\rm km\,s^{-1}}$, both with large uncertainties dominated by distance uncertainty.  J1949+0745 is below the midplane ($z=-0.32_{-0.20}^{+0.15}$ kpc) and moving away from it. It has a well-constrained midplane crossing time of $1.86_{-0.16}^{+0.09}$ Myr. Its likely launch site is in the disk, of order 5 kpc from the Galactic center.

J1251-5059 has a fiducial distance of $d=2.16^{+1.67}_{-1.00}$ kpc, corresponding to $M_{G,0}\approx 7.9$ after extinction correction. This implies a lower luminosity than the three cool \Dsix stars from \citet{Shen2018}, consistent with the expectation from Section~\ref{sec:discoveries} that J1251-5059 is likely the least luminous known member of the cool \Dsix sequence. Its inferred total velocity is $v_{\rm tot}=1563^{+826}_{-364}\,{\rm km\,s^{-1}}$, and its inferred ejection velocity is $v_{\rm ejection}=1508^{+838}_{-413}\,{\rm km\,s^{-1}}$. Unlike J0812-5943 and J1949+0745, J1251-5059 is above the Galactic plane and moving nearly parallel to it. At a distance of $z=0.46_{-0.20}^{+0.34}$ kpc above the midplane, its formal travel time to the disk is $-4.11_{-11.32}^{+2.48}$ Myr. This prevents a reliable kinematic measurement of the star's age.

\begin{figure*}
    \centering
    \includegraphics[width=\textwidth]{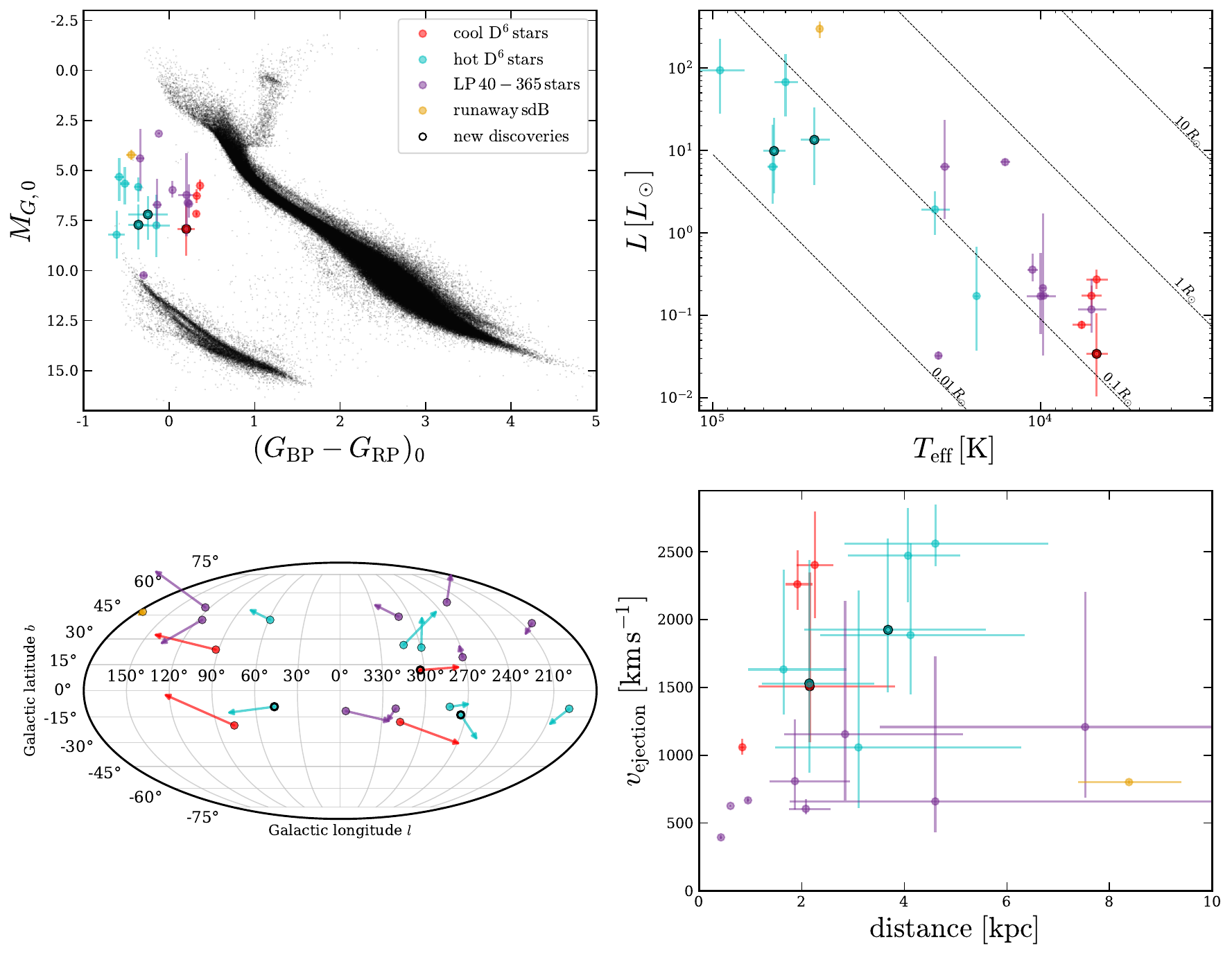}
    \caption{Summary of the known population of hypervelocity WDs and related objects. In each panel, we show suspected hot and cool \Dsix stars in cyan and red, LP 40-365 stars in purple, and the runaway sdB US 708 in yellow. The three new discoveries presented in this work are shown with a black outline. Top left: color-magnitude diagram, with the {\it Gaia} 100 pc sample shown for context. Top right: HR diagram. Bottom left: sky distribution. Arrows show  proper motions, with length scaled to show motion over 0.7 Myr. Bottom right: inferred distances and ejection velocities.}
    \label{fig:populations}
\end{figure*}

\subsection{The known thermonuclear-runaway population}
\label{sec:all_known}

Figure~\ref{fig:populations} shows basic observed and inferred properties of all the objects in Table~\ref{tab:all_systems}. The three new discoveries presented in this work are marked with black outlines. There are now 20 suspected runaways from thermonuclear SNe, including 11 \Dsix stars, 8 LP 40-365 stars, and one runaway sdB. The sample of \Dsix stars includes the ten objects recovered by our search as well as the \Dsix star J1637 discovered by \citet{Raddi2019} and recently characterized by \citet{Hollands2025}. The sample of LP 40-365 stars includes three objects recovered by our search (J1311, J1109, and J1825), as well as LP 40-365 itself \citep{Vennes2017}, J1603 and J0905 \citep{Raddi2019}, J1022 \citep{Bhat2026}, and J1240 \citep{Kepler2016, Gansicke2020}.

We also include US 708, the only confirmed hypervelocity hot subdwarf, in Table~\ref{tab:all_systems} and Figure~\ref{fig:populations}. As described in Appendix~\ref{sec:us708_appendix}, reanalysis of archival and new Keck/ESI spectra gives an updated RV of $835\pm10\,{\rm km\,s^{-1}}$, lower than the value of $917\pm7\,{\rm km\,s^{-1}}$ reported by \citet{Geier2015}. With this RV, we infer an ejection velocity of $802^{+14}_{-13}\,{\rm km\,s^{-1}}$, bringing US 708 in better agreement with expectations for a helium star donor runaway \citep[e.g.][]{Bauer2019, Rajamuthukumar2025}.

In the CMD, all sources except J1240 fall between the WD cooling track and the main sequence, with absolute magnitudes $9 \lesssim M_{G,0} \lesssim 3$ and colors $(G_{\rm BP}-G_{\rm RP})_0 < 0.5$. The sources' blue colors are in part due to selection effects, since our search used an explicit cut of \texttt{bp\_rp} $< 0.5$. However, removing this cut and considering only sources with well-constrained parallaxes still yields only blue sources \citep[e.g.][]{Shen2018}, implying that red hypervelocity runaways are rarer than their blue counterparts at fixed optical luminosity. Consistent with this interpretation, the sources discovered by \citet{Raddi2019} through a color-blind search of SDSS spectra also satisfy  $(G_{\rm BP}-G_{\rm RP})_0 < 0.5$.

The upper right panel of Figure~\ref{fig:populations} shows inferred temperatures and luminosities, with dashed lines showing lines of constant radius. The adopted temperatures are the most recent spectroscopic estimates available in the literature. In cases where no spectroscopic estimate is available (primarily cool \Dsix stars), we estimate temperatures by SED fitting, using the library of C/O-dominated spectral models described by \citetalias{El-Badry2023}. We estimate luminosities by SED fitting, using a spectroscopic temperature prior when spectroscopic temperature estimates exist. We use the distance samples from our kinematic modeling (Section~\ref{sec:orbits}).

The population spans at least a factor of $\sim 20$ in effective temperature and 1000 in luminosity. Most of the observed sources are very likely inflated relative to a mass-radius relation for degenerate objects (although their masses are not well constrained). The only exceptions are J1240, which has $R \approx 0.016 R_\odot$, consistent with predictions for a normal $\approx 0.4\,M_{\odot}$ WD, and perhaps J1332 and J1949, which have poorly constrained distances but could have radii of $\lesssim 0.02\,R_{\odot}$ at their closest possible distances. The hot objects are generally more luminous but smaller than the cool ones. Most of the LP 40-365 stars have temperatures intermediate between hot and cool \Dsix stars, but two hot \Dsix stars -- J1637 and J1235 -- have temperatures overlapping with the LP 40-365 stars.

The temperature distribution of the known \Dsix stars appears bimodal, with one group at $T_{\rm eff}\lesssim 10{,}000\,{\rm K}$, another at $T_{\rm eff}\gtrsim 50{,}000\,{\rm K}$, and comparatively few objects at intermediate temperatures. Because the sample is still small and heterogeneous, we do not attempt to quantify the significance of the bimodality, but we note that it is unlikely to be a product of our sample selection: our color cuts are sensitive to sources with $T_{\rm eff}\gtrsim 6{,}000\,{\rm K}$ and should not preferentially remove objects at intermediate temperatures.


The lower left panel of Figure~\ref{fig:populations} shows the objects' location in the sky. Arrows show proper motions in angular units. Few objects are found at low Galactic latitudes. As expected, most sources are moving away from the disk. However, some objects in all three classes are moving towards the disk. Since all the \Dsix stars and at least some of the LP 40-365 stars are unbound from the Milky Way, this suggests that the objects are relatively young and were launched from a kinematically warm population. Our simulations in Section~\ref{sec:space_density} suggest a progenitor population with a height of $h_z \approx 0.3\,{\rm kpc}$.

The lower right panel shows inferred distances and ejection velocities. Most \Dsix stars have inferred ejection velocities of order $2000\,{\rm km\,s^{-1}}$, while the LP~40-365 stars and US~708 cluster at lower ejection velocities. The two nearest objects are the LP 40-365 stars J1240 and LP 40-365 itself. This suggests that the local space density of LP 40-365 stars is likely higher than that of \Dsix stars. It would be a mistake, however, to conclude that LP 40-365 stars have a higher birth rate than \Dsix stars: LP 40-365 stars remain detectable for longer, since (a) they have typical ejection velocities $\sim$3 times lower than \Dsix stars and thus remain detectable longer, and (b) some LP 40-365 stars remain bound to the Milky Way and thus can remain within a plausibly detectable volume indefinitely. The nearest object, J1240, has an inferred cooling age of $\sim 40$\,Myr \citep{Gansicke2020}, more than an order of magnitude larger than the kinematically-inferred ages of \Dsix stars.

\section{Population modeling of \Dsix stars}
\label{sec:space_density}

Our search for hypervelocity WDs is complete within the selection limits discussed in Section~\ref{sec:candidates}. In total, the query detects 10 of the 11 known  \Dsix stars in Table~\ref{tab:all_systems}.\footnote{The only known \Dsix star not recovered by our search is J1637 \citep{Hollands2025}: our initial {\it Gaia} query does not capture it because at $v_{\perp,\,{\rm lower}}=393\,{\rm km\,s^{-1}}$, it narrowly fails our cut of $v_{\perp,\,{\rm lower}}=400\,{\rm km\,s^{-1}}$. This object was discovered via a template search in SDSS spectra by \citet{Raddi2019}.} To constrain the intrinsic population of \Dsix stars, we perform Monte Carlo simulations of their birth and evolution and compare these to the population recovered by our search.

\subsection{Evolutionary models}
Several works have presented long-term evolutionary models for \Dsix stars and other classes of temporarily inflated SN runaways. All the models we consider are 1D calculations performed with MESA \citep{Paxton2011, Paxton2013, Paxton2015, Paxton2018, Paxton2019, Jermyn2023}; in several cases, the initial conditions are motivated by 3D hydrodynamic simulations. 

The temperatures, luminosities, and evolutionary timescales predicted by these models differ by multiple orders of magnitude. The evolutionary state of the observed stars is not well understood, and so their inferred birth rate depends significantly on the assumed detectable lifetime. We forward-model our survey under the assumption of several different evolutionary models. Comparison of the observed and simulated population constrains the birth rate required to produce the observed population under any one model and can rule out models that make predictions inconsistent with observations. We begin by summarizing the evolutionary models under consideration. 

\subsubsection{WDs shocked by SN ejecta}
\citet{Bhat2025b} modeled the evolution of WDs whose outer layers have been inflated by interaction with a SN shock from an exploding companion. Their initial conditions were based on a hydrodynamic double-detonation simulation from \citet{Pakmor2022}, in which the accreting WD exploded but the donor survived. Only the outer layers of the surviving donor expand, so these models contract and return to the WD cooling sequence within $\lesssim 10^4$ yrs. \citet{Bhat2025b} conclude that these models cannot explain the bulk of the observed population, but propose that they may explain the object J1332. 

\citet{Wong2025} calculated related models for surviving helium WDs shocked by SN ejecta, including the effects of mass loss and subsequent thermal relaxation. These models are useful because they provide a set of shocked-donor tracks with lower bound-remnant masses than the models from \citet{Bhat2025b}. Like the \citet{Bhat2025b} tracks, they heat the survivor relatively superficially compared to the fully convective models discussed below, though the lowest-mass models are heated more deeply. They therefore fade rapidly and are primarily detectable at young ages. Given the low initial donor masses in these models, they are unlikely to produce runaways with velocities larger than $\sim 1600\,{\rm km\,s^{-1}}$.

\subsubsection{Fully convective C/O stars}
\citet{Shen2025} modeled the evolution of hot, fully convective, C/O-dominated stars. He calculated models with masses ranging from 0.1\,$M_\odot$ to 0.5\,$M_{\odot}$, as might be produced if a CO WD is partially disrupted and heated during a violent merger. The models contract along an analog of the \citet{Hayashi1961} track until their cores become radiative, after which they heat up and become more luminous, until they eventually become degenerate and evolve down the WD cooling track. Because the cores -- not just the outer layers -- are heated, the models' evolutionary timescales are much longer than those of \citet{Bhat2025b}. \citet{Shen2025} suggested that the low-mass ($\sim 0.1\,M_{\odot}$) tracks may be suitable for cool \Dsix stars. He also speculated that the higher-mass tracks might be able to explain hot \Dsix stars, but noted that it would be difficult to heat such a massive (and thus, dense) WD enough for it to become fully convective. 

\subsubsection{Partially disrupted runaways from violent mergers}

We consider two models based on simulations of violent mergers in which the donor WD is partially disrupted but a bound remnant survives. One model is taken from \citet{Glanz2025}, who simulated a merger of 0.62 $M_{\odot}$ and 0.69 $M_\odot$ hybrid ``HeCO'' WDs (i.e., CO WDs with unusually thick He envelopes). In this model, a CO-dominated $0.49\,M_{\odot}$ remnant, formed from the paritially disrupted donor, is ejected with a velocity of $2061\,{\rm km\,s^{-1}}$.  Because the donor is partially disrupted during the merger and is closer to the exploding WD than in models where it barely fills its Roche lobe, the companion's ejecta is able to shock and heat its deep interior rather than just its surface. Consequently, the remnant cools significantly more slowly than the intact companions such as modeled in \citet{Bhat2025b}. 

We also consider a model from \citet{Bhat2025}, which is based on the violent merger simulation from \citet{Pakmor2025}. In this calculation, the merger of a 0.7 $M_{\odot}$  and a 1.1 $M_{\odot}$  CO WD results in the ejection of a bound remnant with a mass of only $0.16\,M_\odot$ at a velocity of $2800\,{\rm km\,s^{-1}}$. The remnant is heated during the merger, and \citet{Bhat2025} show that its long-term evolution may produce low-mass, C/O-rich hypervelocity stars with parameters similar to the cool \Dsix stars, J1637, and J1235.

\subsubsection{Partially reheated WDs}

\citet{Zhang2019} produced evolutionary models for partially reheated WDs with a range of masses and heating depths. These models were designed to describe survivors of SNe Iax; in this sense, they are likely more applicable to LP 40-365 stars than to \Dsix stars. However, they provide a useful phenomenological comparison set: they varied both mass and a parameter $f_{\rm env}$, which controls how much of the WD envelope is heated, allowing tracks intermediate between superficial shock heating and the more deeply heated fully convective models. We consider three of these models, all with a mass of $0.3\,M_{\odot}$, but with three different values of $f_{\rm env}$: 0.1, 0.5, and 0.9.

\subsection{Forward-model of the Galactic population}

We assume a constant Galactic birth rate of \Dsix stars, adjusting the rate such that each evolutionary model produces roughly the total number detected by our search (10), or the number of observed runaways that could plausibly be explained by each model. We report this rate relative to the Galactic SNe Ia rate, $\mathcal{R}_{\rm Ia} \approx 5\times 10^{-3} \,{\rm yr^{-1}}$ \citep[e.g.][]{Maoz2014}. We simulate only objects launched within the last 20 Myr, since \Dsix stars travel $\sim 2$ kpc per Myr and thus escape the Galaxy within $\lesssim 10$ Myr.

We sample birth positions from a simple Galactic model consisting of an exponential disk with radial scale length $h_R = 2.6~\mathrm{kpc}$ and vertical scale height $h_z = 0.3~\mathrm{kpc}$, and a spherical bulge component with an exponential density profile of scale length 0.5 kpc \citep[e.g.][]{BlandHawthorn2016}. We assume 5\% of runaways are launched from the bulge. This implies a specific SNe Ia rate that is $\sim 5$ times lower in the bulge than in the disk, reflecting the bulge's old stellar population and the falling delay time distribution of SNe Ia \citep[e.g.][]{Maoz2012, Maoz2014}.

We assume runaways are launched isotropically in the rest frame of the binary. We assume a Gaussian distribution of ejection velocities centered on $2000\,{\rm km\,s^{-1}}$ with a dispersion of $500\,{\rm km\,s^{-1}}$. While this is a simplification, our results are only weakly sensitive to the exact launching velocity assumed, as long as most \Dsix stars have $v_{\rm ejection} \gtrsim 1000\,{\rm km\,s^{-1}}$.

We assign disk progenitors the local circular velocity, while bulge progenitors are assigned an isotropic velocity dispersion of $100\,{\rm km\,s^{-1}}$. We then integrate the launched stars in the same \texttt{MWPotential2014} Galactic potential used for the observed sources using \texttt{galpy} (Section~\ref{sec:orbits}) to obtain their present-day positions and velocities relative to the Sun. We assign temperatures and luminosities  as a function of age using the evolutionary tracks from each model; we calculate absolute magnitudes in the {\it Gaia} bands using bolometric corrections derived from the Kurucz models for C/O dominated atmospheres with $\log g \approx 6$. We assign extinctions to each object based on the 3D dust map of \citet{Wang2025}, assuming a \citet{Cardelli1989} extinction law. For simulated sources beyond the distance limit of the 3D map (typically a few kpc, depending on sightline), we add $A_V=1\,{\rm mag\,kpc^{-1}}$ for any additional path length while the source remains close to the Galactic disk ($|z|<1\,{\rm kpc}$). We predict parallax and proper motion uncertainties in {\it Gaia} DR3 as a function of $G-$band apparent magnitude following the empirical relations reported by \citet{Lindegren2021}, and we perturb the mock-observed astrometry according to these uncertainties. We similarly predict $G_{\rm BP}$ and $G_{\rm RP}$-band flux uncertainties as a function of apparent magnitude in these bands, use these to calculate uncertainties in $G_{\rm BP}-G_{\rm RP}$ colors, and perturb the predicted colors by these uncertainties. Finally, we pass the simulated population through the selection function of our search (Section~\ref{sec:adql}), marking objects as ``detected'' if they pass all the cuts in our {\it Gaia} DR3 ADQL query.

\subsection{Results}

\begin{widetext}
\begin{center}
\captionof{table}{Forward-model results for several classes of evolutionary models.}
\label{tab:model_yields}
\footnotesize
\setlength{\tabcolsep}{2.5pt}
\begin{tabular}{lcccccc}
\hline
Model &
Required birth rate &
$d$ &
$M_{G,0}$ &
$t_{\rm midplane}$ &
$G$ &
$(G_{\rm BP}-G_{\rm RP})_0$ \\
&
(10 detected stars) &
$(\mathrm{kpc})$ &
$(\mathrm{mag})$ &
$(\mathrm{Myr})$ &
$(\mathrm{mag})$ &
$(\mathrm{mag})$ \\
\hline
\multicolumn{7}{l}{\it Shocked donors} \\
$0.64\,M_{\odot}$, \citet{Bhat2025b} &  $0.11 \mathcal{R}_{\rm Ia}$ & $1.38_{-0.41}^{+2.13}$ & $8.98_{-3.17}^{+1.02}$ & $0.76_{-0.57}^{+1.15}$ & $19.57_{-1.14}^{+0.59}$ & $-0.52_{-0.14}^{+0.03}$ \\
$0.85\,M_{\odot}$, \citet{Bhat2025b} &  $0.15 \mathcal{R}_{\rm Ia}$ & $1.41_{-0.54}^{+2.77}$ & $7.19_{-2.49}^{+3.14}$ & $0.47_{-0.34}^{+0.67}$ & $19.15_{-1.12}^{+0.92}$ & $-0.53_{-0.05}^{+0.06}$ \\
$0.95\,M_{\odot}$, \citet{Bhat2025b} &  $0.18 \mathcal{R}_{\rm Ia}$ & $2.07_{-1.04}^{+2.32}$ & $7.43_{-3.08}^{+1.88}$ & $0.42_{-0.45}^{+0.43}$ & $18.44_{-0.38}^{+1.62}$ & $-0.53_{-0.03}^{+0.05}$ \\
$0.45\,M_{\odot}$, \citet{Wong2025} &  $0.11 \mathcal{R}_{\rm Ia}$ & $1.21_{-0.25}^{+0.30}$ & $9.24_{-0.07}^{+0.12}$ & $1.23_{-0.87}^{+1.41}$ & $19.74_{-0.80}^{+0.58}$ & $-0.50_{-0.09}^{+0.05}$ \\
$0.25\,M_{\odot}$, \citet{Wong2025} &  $0.069 \mathcal{R}_{\rm Ia}$ & $1.50_{-0.49}^{+0.69}$ & $8.47_{-0.69}^{+0.30}$ & $1.12_{-0.69}^{+1.29}$ & $19.55_{-0.94}^{+0.71}$ & $-0.38_{-0.10}^{+0.07}$ \\
$0.054\,M_{\odot}$, \citet{Wong2025} &  $0.13 \mathcal{R}_{\rm Ia}$ & $2.28_{-0.80}^{+1.63}$ & $6.78_{-1.16}^{+0.88}$ & $0.45_{-0.27}^{+0.22}$ & $18.98_{-1.24}^{+0.86}$ & $0.20_{-0.11}^{+0.09}$ \\
\hline
\multicolumn{7}{l}{\it Fully convective models (\citealt{Shen2025})} \\
$0.2\,M_{\odot}$ &  $0.032 \mathcal{R}_{\rm Ia}$ & $5.14_{-1.69}^{+1.56}$ & $5.21_{-0.56}^{+0.53}$ & $6.18_{-2.20}^{+2.71}$ & $18.77_{-1.19}^{+1.06}$ & $0.37_{-0.26}^{+0.07}$ \\
$0.3\,M_{\odot}$ &  $0.0057 \mathcal{R}_{\rm Ia}$ & $5.31_{-1.96}^{+2.21}$ & $3.92_{-0.77}^{+1.44}$ & $3.79_{-1.18}^{+2.38}$ & $17.94_{-1.19}^{+1.35}$ & $-0.30_{-0.16}^{+0.31}$ \\
$0.5\,M_{\odot}$ &  $0.0064 \mathcal{R}_{\rm Ia}$ & $4.72_{-2.05}^{+2.57}$ & $5.35_{-3.19}^{+0.84}$ & $1.58_{-0.69}^{+1.04}$ & $18.58_{-2.21}^{+1.21}$ & $-0.51_{-0.07}^{+0.18}$ \\
\hline
\multicolumn{7}{l}{\it Partially reheated models (\citealt{Zhang2019})} \\
Partially reheated $0.30\,M_{\odot}$, $f_{\rm env}=0.1$ &  $0.091 \mathcal{R}_{\rm Ia}$ & $1.37_{-0.38}^{+0.33}$ & $8.83_{-0.21}^{+0.26}$ & $1.11_{-0.61}^{+1.30}$ & $19.54_{-0.60}^{+0.67}$ & $-0.46_{-0.07}^{+0.09}$ \\
Partially reheated $0.30\,M_{\odot}$, $f_{\rm env}=0.5$ &  $0.027 \mathcal{R}_{\rm Ia}$ & $2.29_{-0.89}^{+1.01}$ & $7.67_{-1.36}^{+0.43}$ & $1.31_{-1.09}^{+2.54}$ & $19.70_{-1.49}^{+0.57}$ & $-0.52_{-0.09}^{+0.09}$ \\
Partially reheated $0.30\,M_{\odot}$, $f_{\rm env}=0.9$ &  $0.0044 \mathcal{R}_{\rm Ia}$ & $4.69_{-1.81}^{+2.38}$ & $4.84_{-0.75}^{+1.36}$ & $1.68_{-0.89}^{+1.90}$ & $18.94_{-1.15}^{+0.95}$ & $-0.40_{-0.10}^{+0.11}$ \\
Partially reheated $0.60\,M_{\odot}$, $f_{\rm env}=0.5$ &  $0.030 \mathcal{R}_{\rm Ia}$ & $2.81_{-1.44}^{+2.00}$ & $6.31_{-1.72}^{+1.78}$ & $0.68_{-0.53}^{+1.47}$ & $19.30_{-1.31}^{+0.65}$ & $-0.54_{-0.05}^{+0.05}$ \\
\hline
\multicolumn{7}{l}{\it Violent mergers} \\
Violent merger $0.16\,M_{\odot}$, \citet{Bhat2025} &  $0.0060 \mathcal{R}_{\rm Ia}$ & $4.03_{-1.50}^{+1.71}$ & $5.86_{-0.98}^{+0.90}$ & $1.68_{-0.95}^{+2.58}$ & $19.39_{-1.33}^{+0.74}$ & $-0.20_{-0.08}^{+0.08}$ \\
HeCO merger $0.49\,M_{\odot}$, \citet{Glanz2025} &  $0.059 \mathcal{R}_{\rm Ia}$ & $1.66_{-0.63}^{+2.05}$ & $7.94_{-1.95}^{+1.22}$ & $0.74_{-0.54}^{+1.49}$ & $19.45_{-0.80}^{+0.58}$ & $-0.53_{-0.10}^{+0.05}$ \\
\hline
\multicolumn{7}{l}{\it Observed sources} \\
Observed (all) &  \nodata & $2.20_{-0.44}^{+1.90}$ & $6.72_{-1.02}^{+1.13}$ & $0.64_{-1.70}^{+1.23}$ & $19.22_{-1.44}^{+0.81}$ & $-0.30_{-0.25}^{+0.62}$ \\
Observed (hot only) &  \nodata & $3.88_{-1.83}^{+0.34}$ & $6.51_{-0.92}^{+1.30}$ & $0.87_{-0.64}^{+0.99}$ & $19.40_{-0.34}^{+0.73}$ & $-0.44_{-0.15}^{+0.11}$ \\
Observed (cool only) &  \nodata & $2.03_{-0.68}^{+0.17}$ & $6.72_{-0.71}^{+0.86}$ & $-0.04_{-2.44}^{+1.50}$ & $17.83_{-0.64}^{+1.34}$ & $0.32_{-0.06}^{+0.02}$ \\

\end{tabular}
\par\smallskip
\begin{minipage}{0.90\textwidth}
\footnotesize
\textit{Note.} Comparison of the properties of observed \Dsix stars recovered by our search (bottom) to predictions of several different evolutionary models (Section~\ref{sec:space_density}). For each model, we report the Galactic birth rate required to produce 10 runaways detectable with our search; this is expressed relative to the Galactic SNe Ia rate, $\mathcal{R}_{\rm Ia} = 5\times 10^{-3}\,{\rm yr^{-1}}$. Ranges show the median and middle 68\% of several properties of simulated objects detectable with our search for each model.
\end{minipage}
\end{center}
\end{widetext}

We summarize the results of these simulations in Table~\ref{tab:model_yields} and Figures~\ref{fig:sim_wong2025}--\ref{fig:sim_other_models}. The Galactic birth rate required to produce 10 detectable \Dsix stars varies by a factor of $\sim 40$ across models, with the shocked-donor models requiring birth rates of order $0.1\,\mathcal{R}_{\rm Ia}$ and the more deeply heated models from \citet{Shen2025} requiring birth rates of $\lesssim 0.01\,\mathcal{R}_{\rm Ia}$.  These differences are primarily due to differences in the models' predicted observable lifetimes: models that remain bright in the optical for a long time require a smaller intrinsic rate than models that are faint or short-lived. None of the models we considered imply birth rates close to the SNe Ia rate. The observable properties of detectable \Dsix stars predicted by each model also vary significantly across models, as we discuss below. In each column of Figures~\ref{fig:sim_wong2025}-\ref{fig:sim_other_models}, we compare the ten \Dsix stars detected by our search to a random sample of ten detectable sources simulated with a different evolutionary model. 

\subsubsection{Shocked donor models}
\begin{figure*}[!tbp]
    \centering
    \includegraphics[width=0.90\textwidth]{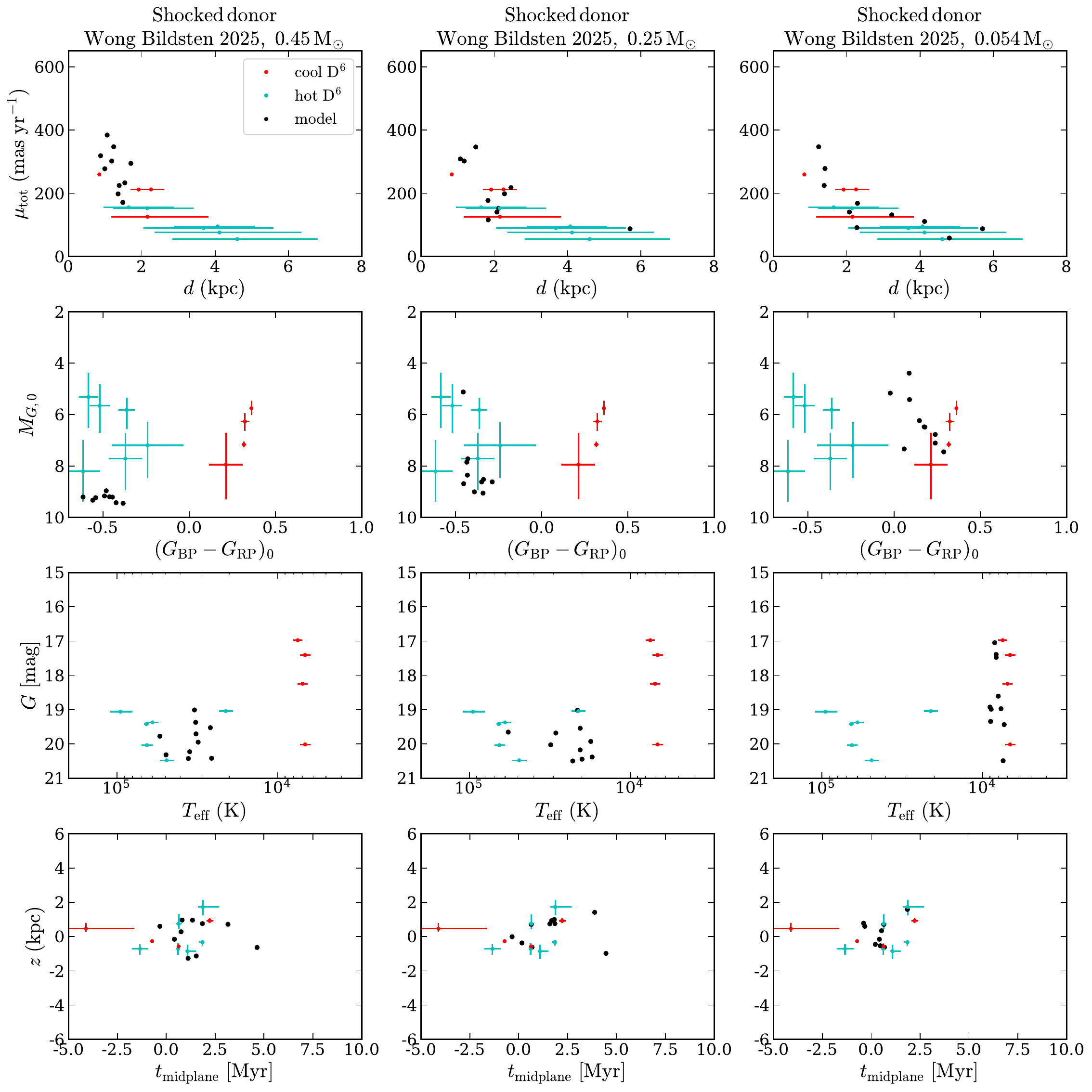}
    \caption{Comparison of observed \Dsix stars recovered by our search (red and cyan) with simulated populations (black) for shocked-donor models from \citet{Wong2025}. Each column shows a different model, labeled by the mass of the runaway star. In this and the following population-comparison figures, the simulated points show a random subset of 10 detectable sources after scaling the model birth rate to produce 10 detections (Table~\ref{tab:model_yields}). Rows show distance and proper motion, intrinsic color and absolute magnitude, spectroscopic temperature and apparent $G$ magnitude, and height above the Galactic plane as a function of midplane-crossing time. These shocked-donor models generally predict sources that are fainter, and therefore detected at smaller distances, than the observed systems. The lowest-mass remnant model produces cooler and redder detections than the more massive models, including some objects with properties similar to the observed cool \Dsix stars.}
    \label{fig:sim_wong2025}
\end{figure*}

Figure~\ref{fig:sim_wong2025} shows predictions for the shocked helium-WD donor models from \citet{Wong2025}. The $0.45\,M_\odot$ and $0.25\,M_\odot$ tracks are generally fainter than the observed \Dsix stars when they satisfy our color and magnitude cuts. The detected simulated sources are therefore typically nearby, with smaller distances and larger proper motions than the bulk of the observed sample.  The lowest-mass model -- which is produced from a $0.10\,M_\odot$ donor that loses significant mass after interaction with the companion's ejecta -- is cooler and redder, and overlaps the observed cool \Dsix stars in the color-magnitude and temperature-magnitude planes. It is likely most applicable to D6-2, whose low velocity is most readily explained by a low-mass donor. 
Producing 10 detections with these models requires birth rates of order $(0.07$--$0.13)\,\mathcal{R}_{\rm Ia}$.

\begin{figure*}[!tbp]
    \centering
    \includegraphics[width=0.90\textwidth]{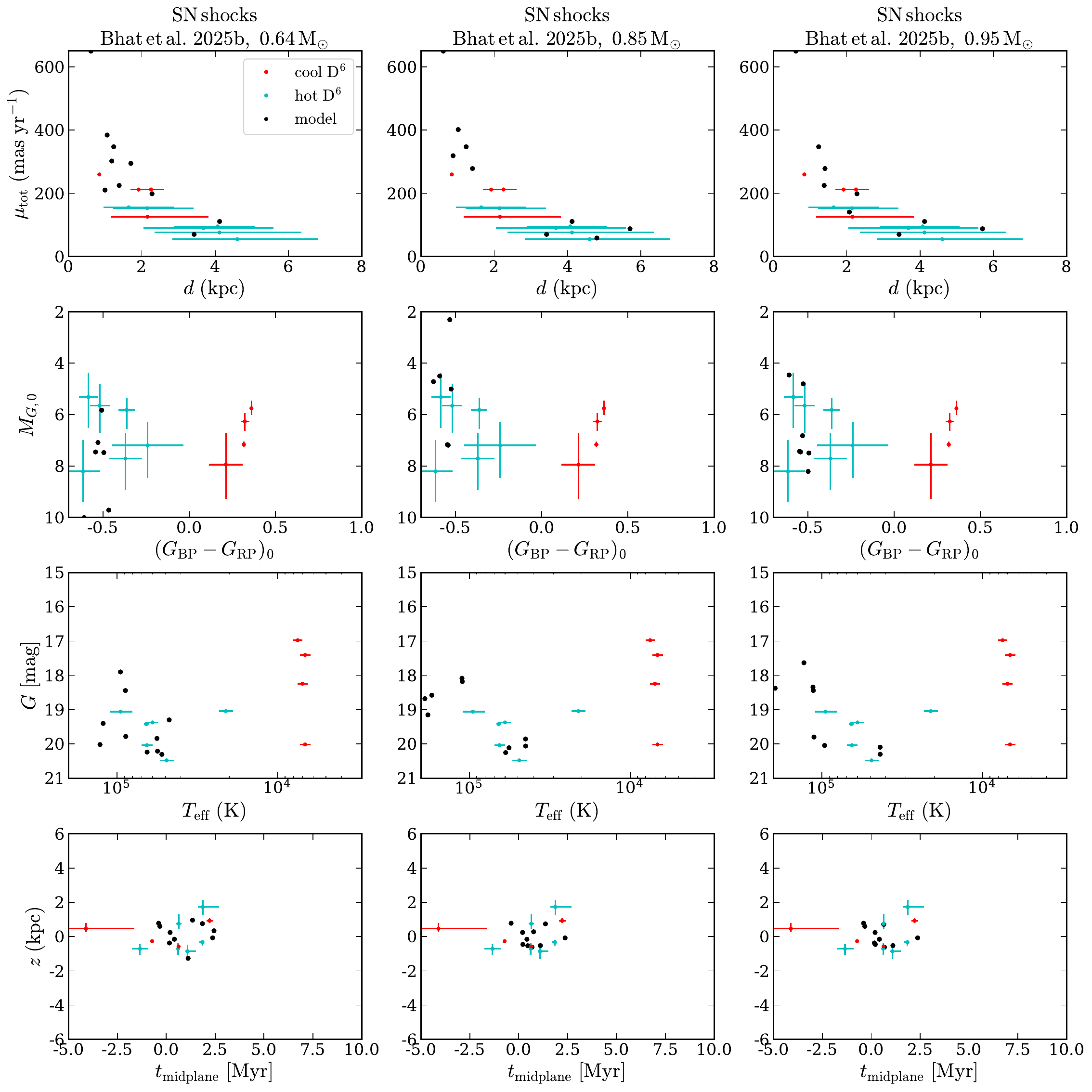}
    \caption{Same as Figure~\ref{fig:sim_wong2025}, but comparing to the shocked-donor models from \citet{Bhat2025b}. The three columns show surviving donor masses of $0.64\,M_{\odot}$, $0.85\,M_{\odot}$, and $0.95\,M_{\odot}$. These models are short-lived because only the outer layers of the surviving donor are inflated by the SN shock. Their detectable phases are hot and blue, but they fade rapidly and therefore require high birth rates to produce 10 detections.}
    \label{fig:sim_bhat2025b}
\end{figure*}

Figure~\ref{fig:sim_bhat2025b} shows the shocked models from \citet{Bhat2025b}, which have higher masses than the models from \citet{Wong2025}. These tracks behave similarly in that they are short-lived, although their detectable phases are hotter and bluer than the \citet{Wong2025} models. They can overlap some observed hot \Dsix stars, especially J1332. However, because the models fade rapidly, producing a sample of 10 detections requires a high birth rate, $\approx 0.1$--$0.2\,\mathcal{R}_{\rm Ia}$, and predicts many nearby, high-proper-motion detections. This makes the shocked-donor models unlikely to explain the full observed population, though they may be appropriate for a few individual systems.

\subsubsection{Fully convective models}
\begin{figure*}[!tbp]
    \centering
    \includegraphics[width=0.90\textwidth]{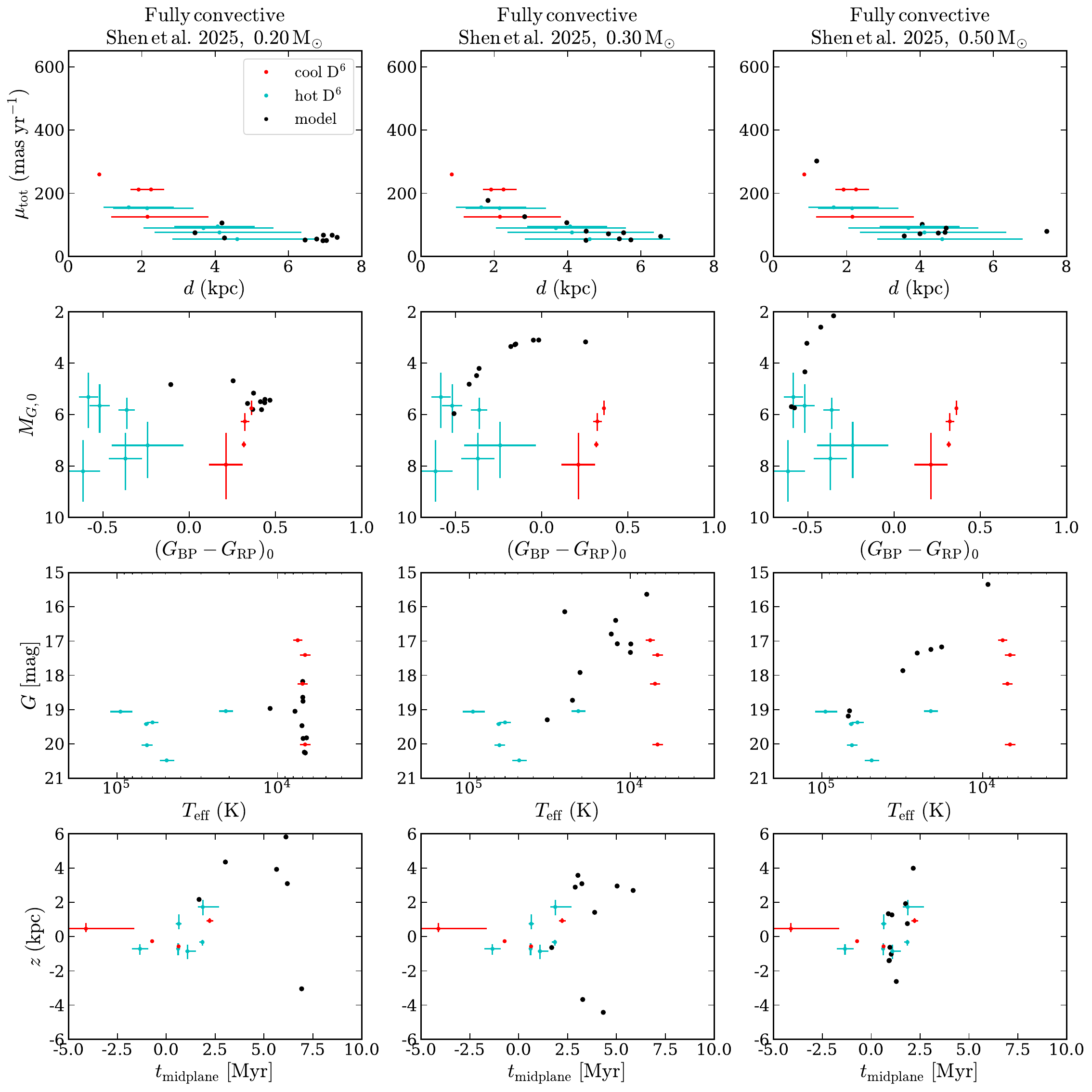}
    \caption{Same as Figure~\ref{fig:sim_wong2025}, but comparing to the fully convective C/O-star models from \citet{Shen2025}. Each column shows a different remnant mass. The $0.2\,M_{\odot}$ and $0.3\,M_{\odot}$ models produce some sources with colors and magnitudes similar to observed cool and hot \Dsix stars, respectively, but predict larger $t_{\rm midplane}$ than most observed systems. The $0.5\,M_{\odot}$ model evolves more quickly and predicts some hot \Dsix-like sources, while also producing detectable sources more luminous than any observed object.}
    \label{fig:sim_shen2025}
\end{figure*}

Figure~\ref{fig:sim_shen2025} shows predictions for three models of different masses from \citet{Shen2025} that begin as fully convective C/O dominated stars. These models are too red to be detected by our search until after they develop radiative cores. The $0.2\,M_{\odot}$, $0.3\,M_{\odot}$, and $0.5\,M_{\odot}$ models reach $G_{\rm BP}-G_{\rm RP} < 0.5$ at ages of 4 Myr, 1.7 Myr, and 0.6 Myr, respectively. The temperature reached during the radiative stage increases with mass, with the three models reaching a maximum $T_{\rm eff}$ of 20\,kK, 36\,kK, and 77\,kK. In the CMD, the $0.2\,M_{\odot}$ model predicts some sources with parameters similar to the observed cool \Dsix stars, while the $0.3\,M_\odot$ and $0.5\,M_{\odot}$ models predict sources resembling some of the hot observed stars. All three models also predict some detectable sources that are more luminous than any of the observed objects.

Because these models only become detectable at relatively late times, they predict $t_{\rm midplane}$ values systematically larger than observed. This tension is modest for the $0.5\,M_{\odot}$ model and relatively severe for the $0.2\,M_\odot$ and $0.3\,M_\odot$ models. It is even more severe for the $0.15\,M_\odot$ model (not shown here), which only becomes blue enough to be detected by our search at an age of 8 Myr. \citet{Shen2025} found that the $0.15\,M_\odot$ model could match the properties of observed cool \Dsix stars at ages of order 1 Myr, in apparent tension with our results. The reason for this tension is that \citet{Shen2025} did not include a cut of $G_{\rm BP}-G_{\rm RP} < 0.5$ when modeling detectability, and the models under consideration are too cool to pass this cut when they are fully convective. That being said, the models predict temperatures only slightly too cool to pass this color cut and enter our search, so changing the assumed opacities or atmosphere models could improve agreement with observation without significantly changing the underlying physical model. If we arbitrarily increase the temperatures of the $0.15\,M_\odot$ track by 500 K, it crosses our color cut after only $\approx 0.2$ Myr rather than $\approx 9$ Myr. With this shift, that model would require a birth rate of $0.017\,\mathcal{R}_{\rm Ia}$ for 10 detections (or $0.007\,\mathcal{R}_{\rm Ia}$ for 4) and would produce a population with colors and magnitudes similar to the observed cool \Dsix stars.

In order to produce 10 detectable sources, the $0.2\,M_\odot$,  $0.3\,M_\odot$, and $0.5\,M_\odot$ models respectively require birth rates of $0.032\,\mathcal{R}_{\rm Ia}$, $0.006\,\mathcal{R}_{\rm Ia}$, and $0.006\,\mathcal{R}_{\rm Ia}$. Since each of these models are likely to be applicable to at most four of the observed objects, the implied fraction of SNe Ia that produce a \Dsix star under these models is $\sim 0.1-1$\%.

\subsubsection{Partially reheated models}
\begin{figure*}[!tbp]
    \centering
    \includegraphics[width=0.90\textwidth]{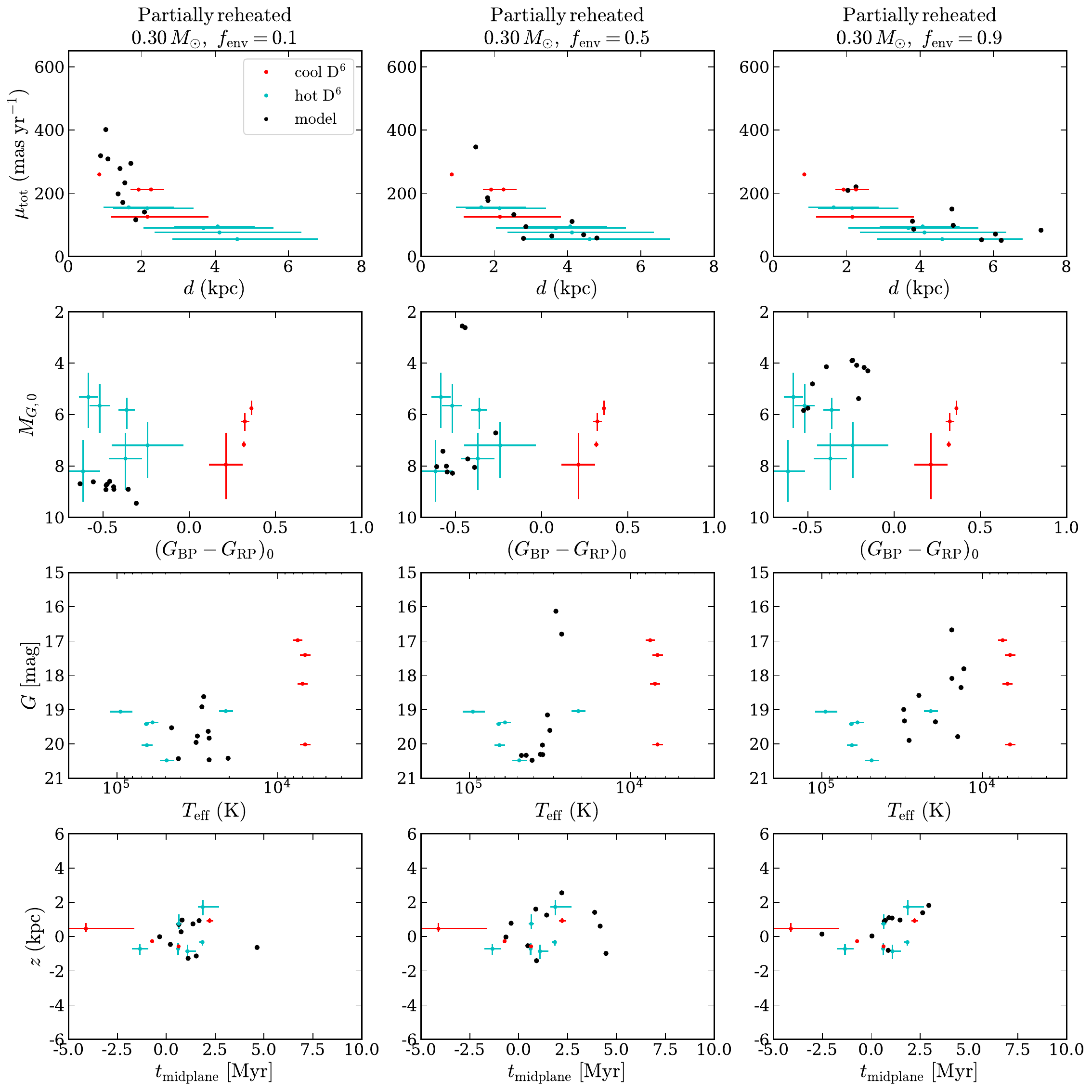}
    \caption{Same as Figure~\ref{fig:sim_wong2025}, but comparing to partially reheated WD models from \citet{Zhang2019}. All three columns show the $0.30\,M_{\odot}$ model, with $f_{\rm env}=0.1$, 0.5, and 0.9 from left to right. Increasing $f_{\rm env}$ heats a larger fraction of the envelope, producing brighter, longer-lived detectable sources and therefore lowering the birth rate required to recover 10 objects.}
    \label{fig:sim_zhang2019}
\end{figure*}

Figure~\ref{fig:sim_zhang2019} shows partially reheated $0.30\,M_{\odot}$ models from \citet{Zhang2019} with three choices of $f_{\rm env}$. This isolates the effect of heating depth at fixed remnant mass. The shallowest-heated model, $f_{\rm env}=0.1$, produces faint detections at close distances and requires a relatively high birth rate, $0.091\,\mathcal{R}_{\rm Ia}$, to yield 10 objects in our search. Increasing the heated envelope fraction makes the detectable phase brighter and longer-lived: the required birth rate falls to $0.027\,\mathcal{R}_{\rm Ia}$ for $f_{\rm env}=0.5$ and $0.0044\,\mathcal{R}_{\rm Ia}$ for $f_{\rm env}=0.9$. The $f_{\rm env}=0.5$ and 0.9 models overlap parts of the observed distribution in color, magnitude, and temperature, but the deepest-heated model tends to predict larger distances and more luminous objects than most of the observed systems. The model with $f_{\rm env} = 0.5$ thus best matches the observed population.

\begin{figure*}[!tbp]
    \centering
    \includegraphics[width=0.90\textwidth]{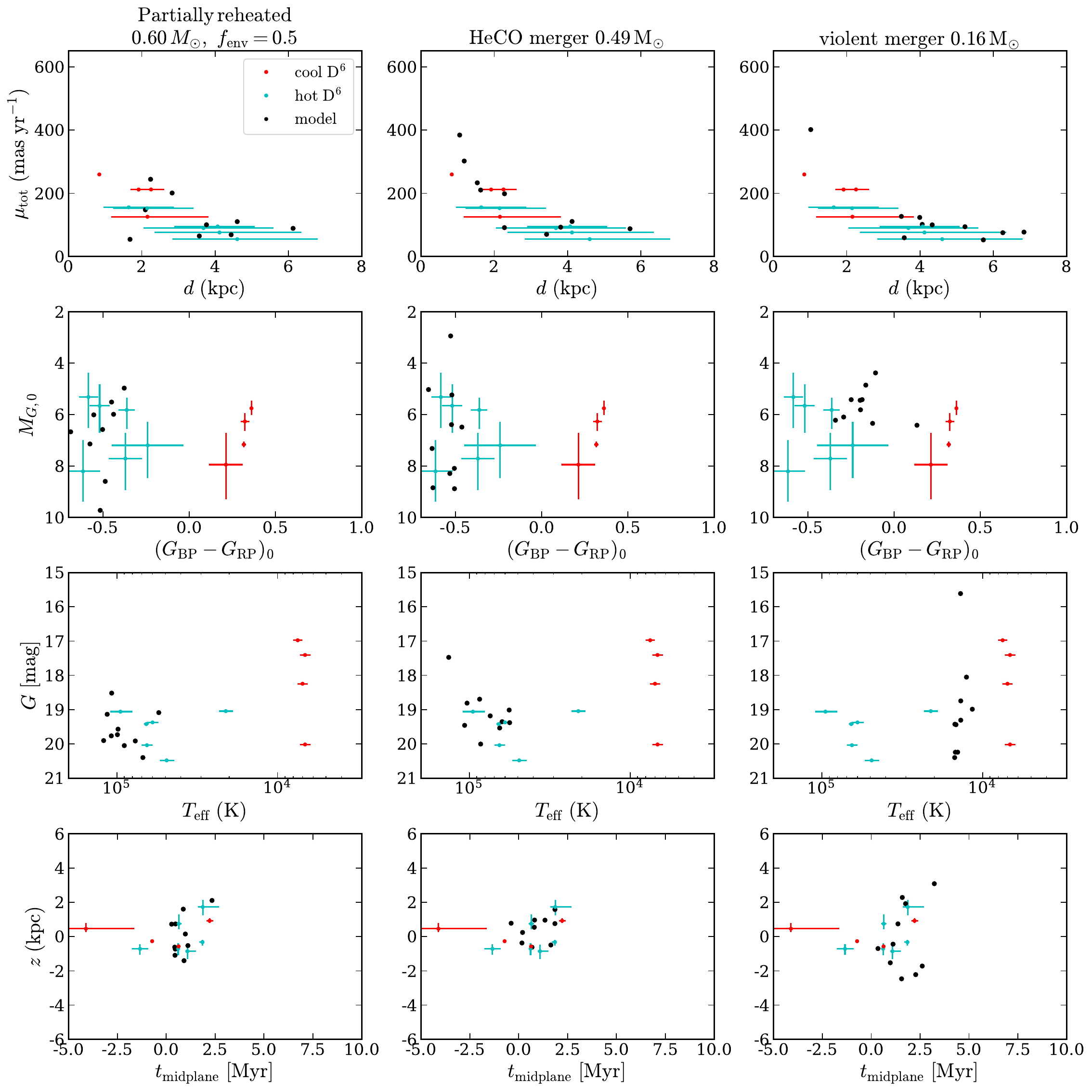}
    \caption{Same as Figure~\ref{fig:sim_wong2025}, but comparing a partially reheated $0.60\,M_{\odot}$ WD model with $f_{\rm env}=0.5$ from \citet{Zhang2019} (left), a $0.49\,M_{\odot}$ WD ejected from a merger of two hybrid WDs (center; \citealt{Glanz2025}), and a $0.16\,M_{\odot}$ remnant ejected from a violent merger (right; \citealt{Bhat2025}). The partially reheated model predicts hot, blue detections but is shorter-lived than the more deeply heated violent-merger remnants. The HeCO model from \citet{Glanz2025} broadly reproduces the observed properties of hot \Dsix stars. The violent-merger model from \citet{Bhat2025} matches some intermediate-temperature systems, with somewhat larger $t_{\rm midplane}$ than the bulk of the observed population.}
    \label{fig:sim_other_models}
\end{figure*}

The left column of Figure~\ref{fig:sim_other_models} shows a partially reheated model with  $f_{\rm env}=0.5$ and a higher mass of $0.60\,M_\odot$. It produces hot, blue detections and requires a birth rate of $0.030\,\mathcal{R}_{\rm Ia}$. Its detections overlap the observed hot \Dsix stars in color and magnitude but predict somewhat higher temperatures than found in the observed population. 

\subsubsection{Violent-merger remnants}
The center and right columns of Figure~\ref{fig:sim_other_models} show violent-merger remnant models. The $0.49\,M_\odot$ HeCO-merger remnant from \citet{Glanz2025} matches the properties of observed hot \Dsix stars well, including their distances, proper motions, luminosities, temperatures, and $t_{\rm midplane}$ distribution. To predict 10 detections, this model requires a birth rate of $0.059\,\mathcal{R}_{\rm Ia}$. The $0.16\,M_{\odot}$ remnant from \citet{Bhat2025} predicts colors and magnitudes intermediate between the observed hot and cool \Dsix stars, and some of the sources predicted to be detectable have physical parameters similar to J1235. Its required birth rate is much lower, about $0.006\,\mathcal{R}_{\rm Ia}$, because the model remains detectable for a relatively long time. These violent-merger calculations effectively produce partial reheating: the survivor is heated more deeply than in the superficially shocked donor models, but it is not reset to the fully convective initial states assumed by \citet{Shen2025}. In this sense, their thermal evolution is qualitatively similar to the intermediate-heating behavior captured phenomenologically by the \citet{Zhang2019} models.

\subsection{Selection effects and future surveys}

\begin{figure*}[!tbp]
    \centering
    \includegraphics[width=0.90\textwidth]{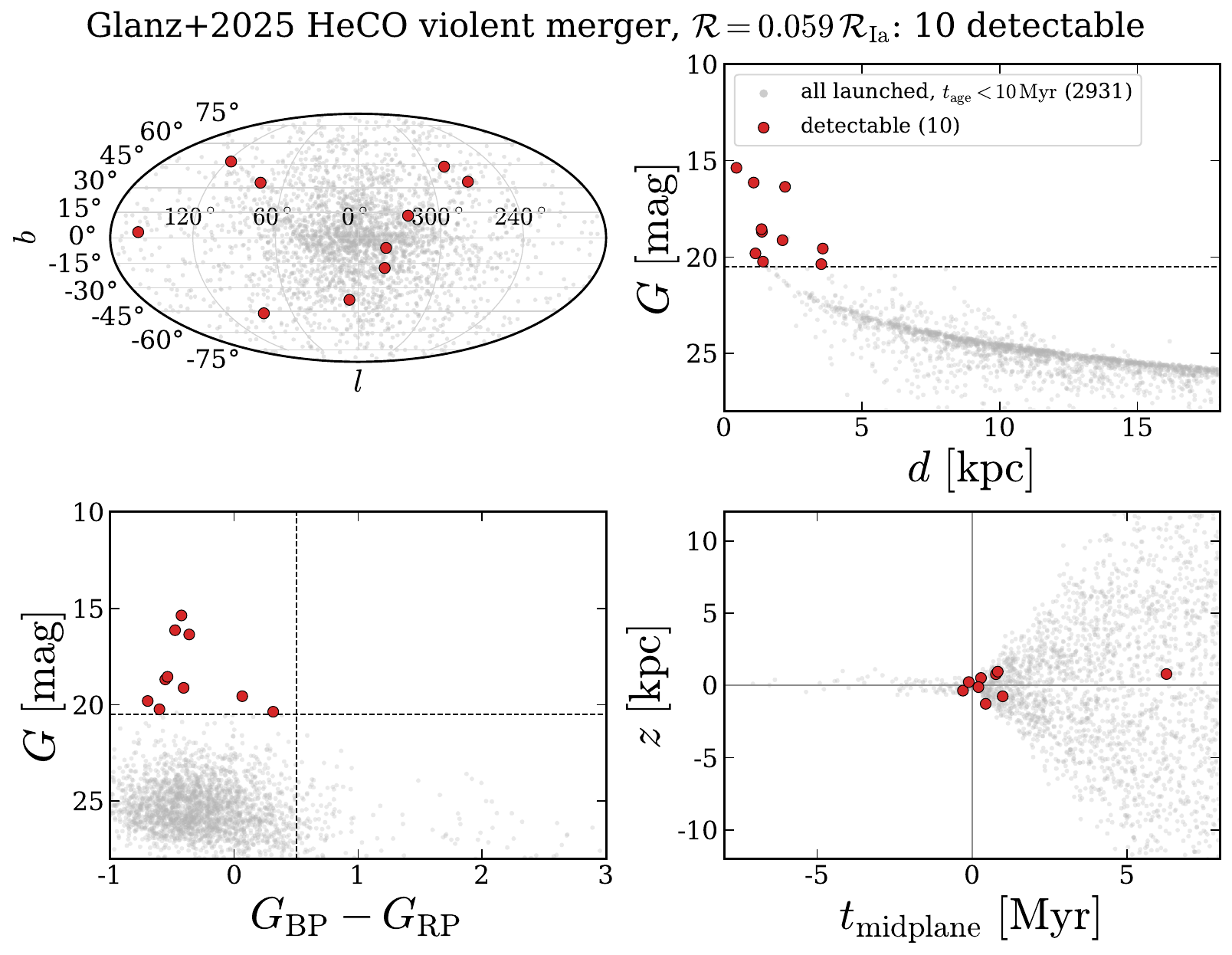}
    \caption{Comparison of the observed and underlying \Dsix star population. Gray points show all simulated runaways launched in the last 10 Myr under the \citet{Glanz2025} evolutionary model, while red points show the subset detectable to our search. The panels show sky position, apparent magnitude as a function of distance and observed color, and height above the Galactic plane as a function of midplane-crossing time. Dashed lines mark the magnitude and color cuts of our search. The model predicts that extending the search to fainter magnitudes would uncover a larger population of \Dsix stars, but such objects could be efficiently selected only with precise astrometry beyond the {\it Gaia} faint limit.  }
    \label{fig:glanz_selection_bias}
\end{figure*}

The birth rates in Table~\ref{tab:model_yields} describe the rate of all runaways in a given evolutionary model, not only the rate of objects that satisfy our search criteria. Figure~\ref{fig:glanz_selection_bias} illustrates this distinction for the HeCO violent-merger model from \citet{Glanz2025}. After re-normalizing this model to produce 10 detections with our current search, it predicts of order 3000 runaways younger than 10 Myr in the Galaxy. Most of these objects are not recovered by our search because they are too faint. 

This modeling provides a simple way to forecast the yield of future searches. Under the \citet{Glanz2025} evolutionary models, extending the magnitude limit of our search from $G<20.5$ to $G<21.5$ (which is fainter than the {\it Gaia} limit and thus currently infeasible) would increase the expected yield from 10 to $\approx 19$ detections. In contrast, relaxing the color cut from $G_{\rm BP}-G_{\rm RP}<0.5$ to $<1.0$ adds only about one object for this model, because the detectable HeCO-merger remnants are already blue. The population predicted to exist at low latitude and reddened beyond our search's color limits is predicted to be modest, since \Dsix stars move $\sim$ 1 kpc away from the Galactic midplane within 0.5 Myr. 
These conclusions are, of course, model dependent: redder evolutionary tracks, such as those predicted for lower-mass models, would benefit more from a relaxed color cut. Similarly, models with short evolutionary lifetimes predict larger gains from searching at low Galactic latitudes, since in those models few objects leave the Galactic plane before fading beyond detectability. 

Future searches will likely struggle to  extend this census to much larger distances. Rubin/LSST and other deep imaging surveys will identify high-proper-motion blue sources below the {\it Gaia} magnitude limit, but without precise parallaxes it will be difficult to distinguish distant hypervelocity candidates from the much larger population of nearby WDs. {\it Gaia} will not provide parallaxes for stars much fainter than those considered here, so our search has likely recovered most of the low-hanging fruit accessible to a simple tangential-velocity selection in the near future. Blind spectroscopic surveys can still expand the sample by detecting objects below the {\it Gaia} limit, or objects above the {\it Gaia} limit whose velocities are primarily radial and therefore missed by our proper-motion selection.

\section{Summary and conclusions}
\label{sec:conclusions}

We have presented a systematic search for hypervelocity WD-like runaways from thermonuclear SNe using {\it Gaia} DR3 astrometry and photometry. Our selection is intentionally simple: we search for blue sources whose {\it Gaia} astrometry implies high tangential velocities. The resulting sample contains 92 candidates. We classify all of them using a combination of new spectroscopy and archival data, giving the search a well-defined selection function suitable for population modeling.

Since the first results of our search were presented by \citetalias{El-Badry2023}, there have been several theoretical developments which shape our interpretation of the observed sources. Following \citet{Bauer2021}, \citetalias{El-Badry2023} interpreted the high velocities of observed \Dsix stars as evidence that they are massive ($M \gtrsim 0.8\,M_{\odot}$), and for this reason, \citetalias{El-Badry2023} concluded that the \citet{Zhang2019} models could not adequately explain the observed population: those models could predict plausible temperatures, luminosities, and ages, but only for lower runaway star masses than implied by the observed ejection velocities.

However, several works have since proposed that observed \Dsix stars are the remnants of donors that were partially disrupted during a WD merger or the resulting explosion \citep[e.g.][]{Shen2025, Pakmor2025, Bhat2025, Glanz2025}. In this case, the surviving remnant can achieve higher velocities than the pre-SN orbital velocity, and there is no simple mapping between ejection velocity and donor mass. In this case, the observed temperatures, luminosities, and ages become easier to explain.

Our main results are summarized below. 

\begin{enumerate}
    \item The search recovers 10 suspected \Dsix stars and three LP~40-365 stars (Table~\ref{tab:survey_accounting}). The confirmed runaways are preferentially brighter, closer to the Galactic plane, and have larger RV amplitudes than the false positives, but they cannot be cleanly selected without spectroscopic follow-up  (Figure~\ref{fig:search_sample}). The false positives separate into a bright population of mostly A/F-type main-sequence stars near the tangential-velocity threshold and a faint population dominated by WDs and other sources with poorly constrained parallaxes.

    \item Three of the suspected \Dsix stars are new discoveries. J1251-5059 is a cool object resembling the original \Dsix sample from  \citet{Shen2018} (Figure~\ref{fig:J1251_spectrum}), while J0812-5943 and J1949+0745 are hot, hydrogen-deficient, C/O rich stars whose spectra resemble the previously known hot \Dsix population (Figures~\ref{fig:J0812_spectrum} and \ref{fig:J1949_spectrum}). Orbit integrations imply ejection velocities above $1000\,{\rm km\,s^{-1}}$ for all three new objects (Figure~\ref{fig:populations}) and slightly fainter absolute magnitudes than the sample of previously known \Dsix stars. Poorly constrained distances limit precise measurements of ejection velocities or luminosities for the three new discoveries. 

    \item The known population of hypervelocity runaways is heterogeneous, spanning at least an order of magnitude in effective temperature and three orders of magnitude in luminosity (Figure~\ref{fig:populations}). The \Dsix stars show tentative evidence for a bimodal temperature distribution (Figure~\ref{fig:populations}), though two objects have intermediate temperatures.  For all the evolutionary models we considered, the effective temperature at late times is primarily a function of mass, with more massive models predicting higher temperatures. The four cool \Dsix stars are best reproduced by low-mass models with $M\lesssim 0.2\,M_\odot$, while the temperature of the hot \Dsix stars appear to imply $M \gtrsim 0.4\,M_\odot$. It is thus natural to ascribe the apparent temperature bimodality to a bimodality in the mass of the surviving remnant.  The inferred birth rate of \Dsix stars varies by more than an order of magnitude across models (Table~\ref{tab:model_yields}), with longer-lived models implying lower birth rates.

    \item No single evolutionary model reproduces the full range of observed \Dsix star properties. Most superficially shocked donor models from \citet{Wong2025} and \citet{Bhat2025b} are on average too faint and short-lived to explain the full sample (Figures~\ref{fig:sim_wong2025} and \ref{fig:sim_bhat2025b}), but the lowest-mass models from \citet{Wong2025} are a reasonably good match to the cool \Dsix stars, especially D6-2. Fully convective models from \citet{Shen2025} can produce long-lived, more luminous remnants, but these models generally predict older ages at the time of detection than inferred for the observed population (Figure~\ref{fig:sim_shen2025}). This tension could be alleviated by modest increases in the temperature of these models. Partially reheated models from \citet{Zhang2019} span a useful range of behavior, with hotter and longer-lived detectable populations as a larger fraction of the envelope is reheated (Figure~\ref{fig:sim_zhang2019}).  The observed population is difficult to reconcile with a single-parameter family of models. The data instead favor a heterogeneous population in which both remnant mass and heating depth vary substantially across systems.

    \item Of the evolutionary models we considered, violent-merger remnants and other partially reheated models match the ages, colors, and magnitudes of the largest number of observed \Dsix stars (Figure~\ref{fig:sim_other_models}). These models predict runaway stars that are heated more deeply than superficially shock-heated models but that still fade more rapidly than fully convective models. None of the models fully explain the cool \Dsix star population, as the range of ejection velocities across the four observed objects likely requires multiple formation pathways. The low-mass models from \citet{Shen2025} and \citet{Bhat2025} -- which are most simply understood as being products of violent mergers -- may produce objects similar to D6-1, D6-3, and J1251-5059. The low-mass shocked donor models from \citet{Wong2025} seem most promising to explain the physical parameters and low velocity of D6-2. 

    The ``hybrid'' HeCO merger model from \citet{Glanz2025} predicts a detected population similar to the observed hot \Dsix stars. This does not necessarily imply that this specific model is the most correct. Rather, it suggests that the degree and depth of heating in that calculation is closer to the effective thermal evolution of the observed remnants -- particularly the hot remnants -- than either superficially shocked or fully reheated models.  

    \item For the models that best resemble the observed sample, the required birth rates are a few percent of the Galactic SNe Ia rate  (Table~\ref{tab:model_yields}). This suggests that most SNe Ia do not produce an observable \Dsix-like remnant in the region of parameter space probed by our search. This conclusion applies only to detectable remnants; we cannot rule out a scenario in which many SNe Ia produce surviving remnants that are fainter than the currently-known classes of runaways. The simplest explanation for the low inferred birth rate of SN runaways is that most Type Ia SNe cause both WDs to explode, as has been predicted by several recent calculations \citep[e.g.][]{Pakmor2022, Boos2024, Shen2024, Pollin2024}.

    Another possible explanation is that most SNe Ia arise from a single-degenerate channel, which would produce lower-velocity runaways not detectable with our search \citep[e.g.][]{Marietta2000, Pan2012, Rau2022}. However, this scenario struggles to explain the lack of detected runaways in remnants of historical supernovae \citep{Kerzendorf2012, Kerzendorf2013, Kerzendorf2014, Kerzendorf2018, Shields2023}, as well as the delay time distribution and several other population properties of SNe Ia \citep[e.g.][]{Maoz2014}. Other channels in which no companion star survives \citep[e.g.][]{Soker2026} are also consistent with the results of our search.
    
\end{enumerate}

\section*{acknowledgments}
We thank Lars Bildsten, Evan Bauer, Ruediger Pakmor, Jim Fuller, Logan Proust, Abinaya Rajamuthukumar, and Stephan Geier for useful discussion related to this work.

This work was supported by NSF grants AST-2508988 and AST-2205631, NASA/ESA
Hubble Space Telescope program No.\ 
17441, and Scialog grant \#SA-LSST-2024-114c from the Research Corporation for Science Advancement. The Kavli Institute for Theoretical Physics (KITP) hosted the program, ``White Dwarfs as Probes of the Evolution of Planets, Stars, the Milky Way, and the Expanding Universe,'' during which this project was initiated. This research was supported in part by the U.S. National Science Foundation (NSF) under grants PHY-1748958. This research benefited from discussions that were
funded by the Gordon and Betty Moore Foundation
through Grant GBMF5076.

We thank the staffs of the various observatories at which data were obtained. This work is partially based on observations obtained at the Southern Astrophysical Research (SOAR) telescope, which is a joint project of the Minist\'{e}rio da Ci\^{e}ncia, Tecnologia e Inova\c{c}\~{o}es (MCTI/LNA) do Brasil, the US National Science Foundation’s NOIRLab, the University of North Carolina at Chapel Hill (UNC), and Michigan State University (MSU). Some of the data presented herein were obtained at the W. M. Keck Observatory, which is operated as a scientific partnership among the California Institute of Technology, the University of California, and NASA; the observatory was made possible by the generous financial support of the W. M. Keck Foundation.

This research has made use of the Keck Observatory Archive (KOA), which is operated by the W. M. Keck Observatory and the NASA Exoplanet Science Institute (NExScI), under contract with the National Aeronautics and Space Administration. 

 This work has made use of data from the European Space Agency (ESA) mission {\it Gaia} (\url{https://www.cosmos.esa.int/gaia}), processed by the {\it Gaia} Data Processing and Analysis Consortium (DPAC, \url{https://www.cosmos.esa.int/web/gaia/dpac/consortium}). Funding for the DPAC has been provided by national institutions, in particular the institutions participating in the {\it Gaia} Multilateral Agreement.

\newpage

\bibliographystyle{mnras}

\appendix

\section{An updated radial velocity for US 708}
\label{sec:us708_appendix}

US 708 is the only confirmed hypervelocity hot subdwarf. \citet{Geier2015} reported an RV of $917\pm 7\,{\rm km\,s^{-1}}$ for US 708 based on Keck/ESI spectra. That RV led \citetalias{El-Badry2023} to infer an ejection velocity of $833^{+17}_{-16}\,{\rm km\,s^{-1}}$, which is uncomfortably high for the supernova runaway scenario with a helium-star donor \citep{Justham2009, Wang2009, Neunteufel2020, Neunteufel2022}, though it could be explained if the progenitor binary was on a halo orbit \citep[e.g.][]{Brown2015b}.

We observed US 708 with Keck/ESI on MJD 60343 and measured an RV of $818 \pm 15\,{\rm km\,s^{-1}}$. This value is $\approx 100\,{\rm km\,s^{-1}}$ slower than that measured by \citet{Geier2015}, but still $\approx 100\,{\rm km\,s^{-1}}$ faster than the RV of $708\pm 15\,{\rm km\,s^{-1}}$ originally reported by \citet{Hirsch2005}. To understand the origin of this tension, we retrieved the 10 ESI spectra of US 708 presented by \citet{Geier2015} from the Keck archive and re-reduced them using \texttt{MAKEE}. From the coadded spectrum, we measure an RV of $835\pm 10\,{\rm km\,s^{-1}}$, which is consistent with the measurement from our data and inconsistent with the RV reported by \citet{Geier2015}. Additional unpublished spectra confirm this lower RV (S. Geier, private communication). 

We conclude that the true RV of US 708 is $835\pm 10\,{\rm km\,s^{-1}}$, about $80\,{\rm km\,s^{-1}}$ lower than assumed in most analyses in the last decade. We suspect that the different RV reported by \citet{Geier2015} is due to an error in air-to-vacuum wavelength conversion. Integrating US 708's trajectory with this new RV yields an ejection velocity of $802^{+14}_{-13}\,{\rm km\,s^{-1}}$. This is only $\sim 30\,{\rm km\,s^{-1}}$ slower than reported by \citetalias{El-Badry2023}; our lower RV is partially offset by the fact that we integrate orbits in a realistic Galactic potential rather than assuming straight-line trajectories. Our inferred ejection velocity is in plausible agreement with predictions of simulations for sdB + WD binaries in which the WD undergoes a double detonation after accreting helium from an sdB star, though still on the high side for sdB stars massive enough to continue burning helium \citep{Bauer2019, Rajamuthukumar2025}.

\section{Summary of follow-up}
\label{sec:appendix}

\setlength{\LTleft}{0pt}
\setlength{\LTright}{0pt}

\begingroup
\footnotesize
\renewcommand{\arraystretch}{1.42}
\setlength{\extrarowheight}{1.5pt}
\setlength\tabcolsep{0.09cm}
\setlength{\LTcapwidth}{\textwidth}
\makeatletter
\def\LT@makecaption#1#2#3{%
  \LT@mcol\LT@cols l{%
    \parbox[t]\LTcapwidth{%
      \reset@font
      #1{#2: }#3%
      \endgraf\vskip\baselineskip
    }%
  }%
}
\makeatother
\begin{longtable}{@{}>{\strut}llcccccccl@{}}
\caption{All {\it Gaia} sources returned by the ADQL query from Section~\ref{sec:adql}, together with results from our follow-up spectroscopy and archival classifications. Sources are sorted by $v_{\perp,{\rm lower}}=4.74\mu/(\varpi+\sigma_\varpi)$, the $1\sigma$ lower limit on their tangential velocity. $v_{\perp}$ is a point-estimate of the sources' tangential velocity (Equation~\ref{vperp}). Both $v_{\perp}$ and $v_{\perp,{\rm lower}}$ can be negative for sources with negative parallaxes; this implies a large but poorly constrained tangential velocity.  Objects with common names are detections recovered by the search, while unnamed sources are false positives. References: S18: \citet{Shen2018}; R19: \citet{Raddi2019}.}
\label{tab:query}\\
Source ID & Name & $G$  & $\varpi$ & $\mu$ & $v_\perp$ & $v_{\perp,{\rm lower}}$ & RV & verdict & follow-up \\
 &  & [mag]  & [mas] & [$\rm mas\,yr^{-1}$] & [$\rm km\,s^{-1}$] & [$\rm km\,s^{-1}$] & [$\rm km\,s^{-1}$] &  &  \\
\hline
\endfirsthead

\multicolumn{10}{l}{{\tablename\ \thetable{} -- continued}} \\
Source ID & Name & $G$  & $\varpi$ & $\mu$ & $v_\perp$ & $v_{\perp,{\rm lower}}$ & RV & verdict & follow-up \\
 &  & [mag]  & [mas] & [$\rm mas\,yr^{-1}$] & [$\rm km\,s^{-1}$] & [$\rm km\,s^{-1}$] & [$\rm km\,s^{-1}$] &  &  \\
\hline
\endhead
2448037457850504832 &   &  20.40 & $-0.85 \pm 0.91$ & 122.8 & -687 & 9167 &   & \red{DA WD} & SOAR \\
2402028733088281984 &   &  20.28 & $-0.77 \pm 0.80$ & 59.0 & -362 & 9145 & $-10\pm 50$ & \red{MS} & SOAR \\
5290552672404889984 & J0812-5943  &  20.48 & $-0.51 \pm 0.57$ & 89.9 & -833 & 6841 & $-1200\pm 10$ & \green{hot \Dsix}  & SOAR/GMOS \\
1357638170128486400 &   &  20.48 & $-0.62 \pm 0.71$ & 63.5 & -481 & 3517 &   & \red{blended} &  \\
6156470924553703552 &  J1235-3752 &  19.05 & $-0.10 \pm 0.24$ & 95.2 & -4685 & 3183 & $-1694\pm 10$ & \green{hot \Dsix} & MagE \\
2156908318076164224 &  D6-3  &  18.25 & $0.42 \pm 0.10$ & 212.0 & 2374 & 1922 & $-20\pm 80$ & \green{cool \Dsix} & S18 \\
6736291479453047552 &   &  20.17 & $-0.71 \pm 0.86$ & 52.5 & -349 & 1704 & $-120\pm 100$ & \red{DA WD} & SOAR/GMOS \\
5805243926609660032 & D6-1  &  17.41 & $0.53 \pm 0.07$ & 211.7 & 1890 & 1669 & $1200 \pm 40$ & \green{cool \Dsix} & S18 \\
4080377230814011136 &   &  20.34 & $-1.04 \pm 1.22$ & 55.6 & -253 & 1485 &  & \red{blended} &  \\
3917003370724484352 &   &  20.44 & $-0.91 \pm 1.28$ & 100.7 & -523 & 1285 &   & \red{DC WD} & SOAR/LRIS \\
6003972342578025344 &   &  20.10 & $-0.60 \pm 0.83$ & 52.4 & -411 & 1101 &  & \red{blended} &  \\
1356636716899212800 &   &  20.33 & $-0.23 \pm 0.53$ & 66.9 & -1388 & 1056 & $-160\pm 50$ & \red{DA WD} & LRIS \\
4184081105961170816 &   &  20.43 & $-0.42 \pm 0.76$ & 74.7 & -843 & 1030 & $-60\pm 50$ & \red{MS} & LRIS \\
3992451210082879488 &   &  20.16 & $-0.45 \pm 0.87$ & 86.7 & -907 & 994 & $180\pm 50$ & \red{DA WD} & LRIS \\
1798008584396457088 & D6-2  &  16.98 & $1.19 \pm 0.07$ & 259.5 & 1030 & 977 & $80\pm 10$ & \green{cool \Dsix} & S18 \\
3335306915849417984 & J0546+0836  &  19.06 & $0.07 \pm 0.31$ & 76.1 & 5289 & 942 & $1200\pm 20$ & \green{hot \Dsix} & LRIS/ESI/X-shooter \\
4298029440184716160 & J1949+0745  &  20.04 & $0.20 \pm 0.59$ & 152.8 & 3599 & 911 & $100 \pm 40$ & \green{hot \Dsix} & LRIS \\
468855092650116480 &   &  20.42 & $0.01 \pm 0.76$ & 128.1 & 43058 & 783 &  & \red{blended} &  \\
1259197996445745792 &   &  20.23 & $-0.10 \pm 0.56$ & 75.2 & -3559 & 774 & $-40\pm 50$ & \red{DA WD} & LRIS \\
5250394728194220800 & J0927-6335  &  19.37 & $0.13 \pm 0.21$ & 54.9 & 2062 & 764 & $-2285 \pm 20$ & \green{hot \Dsix} & MagE/X-shooter \\
6712254192464970368 &   &  12.64 & $0.32 \pm 0.02$ & 53.9 & 794 & 744 &  & \red{MS} & GaiaXP \\
4396371856111979776 &   &  20.34 & $-0.64 \pm 0.97$ & 51.6 & -380 & 741 & $-450\pm 100$ & \red{DA WD} & SOAR \\
1153320661688203136 &   &  20.10 & $0.01 \pm 0.64$ & 101.1 & 34551 & 728 & $180\pm 50$ & \red{DA WD} & SOAR \\
2289780724881070592 &   &  20.48 & $-0.09 \pm 0.53$ & 65.4 & -3413 & 702 & $-100\pm 50$ & \red{DA WD} & LRIS \\
6621783871766231296 &   &  20.13 & $-0.38 \pm 0.73$ & 51.0 & -632 & 698 &   & \red{blended} &  \\
1044958121413909888 &   &  20.45 & $-0.12 \pm 0.72$ & 87.5 & -3386 & 693 & $260\pm 50$ & \red{DA} & LRIS \\
1160986392332702720 &   &  15.14 & $0.41 \pm 0.03$ & 62.8 & 730 & 677 & $-117 \pm 20$ & \red{MS} & SDSS \\
3946876384391994496 &   &  14.28 & $0.40 \pm 0.02$ & 59.2 & 700 & 662 & $76\pm 12$ & \red{MS} & \citet{Brown2008} \\
6727110900983876096 &  J1825-3757 &  13.26 & $1.05 \pm 0.03$ & 147.8 & 666 & 649 & $-47\pm 5$ & \green{LP 40-365} & R19 \\
6563500886388740864 &   &  20.15 & $0.19 \pm 0.48$ & 90.8 & 2320 & 644 & $-50\pm 50$ & \red{DA WD} & SOAR \\
1473381658945235584 &   &  20.30 & $0.51 \pm 0.53$ & 136.5 & 1266 & 621 & $30 \pm 50$ & \red{DA WD} & LRIS \\
3688712561723372672 &   &  18.77 & $1.42 \pm 0.20$ & 211.6 & 706 & 620 & $140\pm 14$ & \red{DA WD} & SDSS \\
3983606360591141504 &   &  14.02 & $0.87 \pm 0.02$ & 116.2 & 634 & 620 & $122\pm 10$ & \red{MS} & LAMOST \\
6078434976557457536 & J1251-5059  &  20.02 & $0.43 \pm 0.53$ & 125.7 & 1380 & 619 & $-870\pm 5$ & \green{cool \Dsix} & SOAR/GMOS/X-shooter \\
6775068295327939456 &   &  20.45 & $-0.11 \pm 0.65$ & 69.8 & -2932 & 616 & $-120\pm 50$ & \red{DA WD} & SOAR \\
6164642052589392512 &  J1332-3541 &  19.42 & $0.66 \pm 0.54$ & 155.5 & 1112 & 613 & $1090\pm 50$ & \green{hot \Dsix} & MagE/LRIS \\
4201939094650159488 &   &  20.12 & $-0.52 \pm 1.02$ & 64.8 & -590 & 613 &  & \red{blended} &   \\
724149735422162816 &   &  10.77 & $1.04 \pm 0.02$ & 136.0 & 619 & 606 & $92\pm 10$ & \red{MS} & LAMOST \\
3542263595793124480 &   &  19.18 & $0.67 \pm 0.46$ & 144.3 & 1016 & 604 &  & \red{blended} &  \\
6194859965015895808 &   &  14.56 & $0.41 \pm 0.03$ & 55.7 & 650 & 602 &  & \red{MS} & GaiaXP \\
4709163672760735488 &   &  11.57 & $0.56 \pm 0.02$ & 73.5 & 617 & 600 & $-100\pm 3$ & \red{MS} & GaiaRVS \\
2316981409896303232 &   &  15.23 & $0.56 \pm 0.03$ & 75.8 & 636 & 599 & $50 \pm 50$ & \red{MS} & SOAR \\
3637020190774216064 &   &  20.46 & $0.07 \pm 1.03$ & 138.4 & 9234 & 594 & $40\pm 50$ & \red{DA WD} & SOAR \\
3804182280735442560 & J1109+0001  &  19.09 & $0.40 \pm 0.31$ & 87.6 & 1051 & 592 & $100\pm 10$ & \green{LP 40-365} & LRIS/X-shooter \\
5814962273679342208 &   &  14.79 & $0.40 \pm 0.02$ & 52.7 & 621 & 588 &   & \red{MS} & GaiaXP \\
300228113890870528 &   &  14.95 & $0.40 \pm 0.03$ & 53.4 & 636 & 588 &   & \red{MS} & GaiaXP \\
5703888058542880896 &   &  19.60 & $1.36 \pm 0.32$ & 207.9 & 723 & 586 & $270\pm 50$ & \red{DA WD} & LRIS \\
5517276097516408576 &   &  19.33 & $0.10 \pm 0.35$ & 55.3 & 2735 & 582 &  & \red{blended} &  \\
6610315175214347264 &   &  14.43 & $0.41 \pm 0.02$ & 52.4 & 610 & 578 &  & \red{MS} & GaiaXP \\
3792840680855962240 &   &  15.07 & $0.46 \pm 0.03$ & 60.1 & 619 & 576 & $290\pm 10$ & \red{MS} & LAMOST \\
4771417432717575680 &   &  19.93 & $0.13 \pm 0.32$ & 54.2 & 1998 & 571 & $120\pm 50$ & \red{DA WD} & SOAR \\
660204643116592768 &   &  20.38 & $0.72 \pm 0.66$ & 165.3 & 1082 & 564 & $-100 \pm 50$ & \red{DA WD} & LRIS \\
4998384365291044096 &   &  15.00 & $0.45 \pm 0.04$ & 58.0 & 606 & 560 &  & \red{MS} & GaiaXP \\
5474401427567487616 &   &  20.40 & $-0.28 \pm 0.73$ & 52.8 & -881 & 559 & $350\pm 50$ & \red{DA WD} & SOAR \\
6021110567768770816 &   &  20.26 & $-0.21 \pm 0.78$ & 67.4 & -1512 & 557 & $140\pm 50$ & \red{blended} & SOAR \\
3739128262233092352 &   &  20.21 & $-0.04 \pm 0.58$ & 62.6 & -7973 & 550 & $60\pm 50$ & \red{DB WD} & DBSP,ESI \\
6777304839418469248 &   &  20.44 & $-0.22 \pm 0.77$ & 61.7 & -1308 & 534 & $-180\pm 100$ & \red{DA WD} & SOAR \\
4546525523392712064 &   &  19.39 & $0.52 \pm 0.32$ & 93.6 & 860 & 534 & $-240\pm 50$ & \red{DA WD} & DBSP \\
6853349473073333632 &   &  19.56 & $0.50 \pm 0.46$ & 107.7 & 1027 & 532 & $90 \pm 50$ & \red{DA WD} & SOAR/GMOS \\
3002601569231013504 &   &  20.12 & $0.53 \pm 0.65$ & 130.8 & 1180 & 526 & $180\pm 50$ & \red{DA WD} & SOAR \\
2393804867149529856 &   &  19.78 & $0.13 \pm 0.41$ & 58.8 & 2167 & 513 & $-40\pm 50$ & \red{DA WD} & SOAR \\
3507697866498687232 & J1311-1846 &  18.26 & $0.60 \pm 0.19$ & 83.1 & 653 & 496 & $55\pm 10$ & \green{LP 40-365} & LRIS/X-shooter \\
5361430834062743424 &   &  20.13 & $0.36 \pm 0.52$ & 90.6 & 1186 & 484 & $-230\pm 50$ & \red{DA WD} & SOAR \\
6508240879378242560 &   &  20.11 & $0.37 \pm 0.59$ & 96.0 & 1218 & 473 &   & \red{blended} &  \\
5998866829060560768 &   &  19.49 & $0.73 \pm 0.44$ & 115.6 & 754 & 468 & $-65\pm 10$ & \red{MS} & MagE \\
5183592902806605824 &   &  19.81 & $0.73 \pm 0.58$ & 128.2 & 836 & 463 & $-20\pm 50$ & \red{sdO/B} & SOAR \\
4781587644689625600 &   &  20.28 & $0.14 \pm 0.57$ & 69.4 & 2347 & 462 & $160 \pm 50$ & \red{DA WD} & SOAR \\
4129413800145771776 &   &  19.30 & $0.65 \pm 0.35$ & 96.2 & 697 & 452 & $-290\pm 80$ & \red{DA WD} & LRIS \\
1512757030058082304 &   &  19.94 & $0.35 \pm 0.33$ & 64.3 & 879 & 450 & $-70\pm 50$ & \red{DA WD} & LRIS \\
1271663056690275712 &   &  19.20 & $0.46 \pm 0.22$ & 65.0 & 669 & 450 & $-118\pm 38$ & \red{DA WD} & SDSS \\
1533746771456004480 &   &  20.04 & $0.71 \pm 0.58$ & 121.4 & 813 & 447 & $-194 \pm 43$ & \red{DA WD} & SDSS \\
6536783647184624640 &   &  19.94 & $0.07 \pm 0.53$ & 55.9 & 3844 & 444 & $70\pm 50$ & \red{DA WD} & SOAR/GMOS \\
4096984529352659584 &   &  19.60 & $0.40 \pm 0.49$ & 83.0 & 983 & 440 & $60\pm 50$ & \red{MS} & LRIS \\
6592388973858801920 &   &  18.45 & $0.39 \pm 0.19$ & 52.3 & 638 & 425 & $170\pm 40$ & \red{DA WD} & SOAR \\
5772035376916929408 &   &  20.17 & $0.17 \pm 0.43$ & 51.8 & 1478 & 415 & $190 \pm 50$ & \red{DA WD} & SOAR \\
6729448153444659712 &   &  20.45 & $0.87 \pm 1.00$ & 161.6 & 885 & 410 & $-60\pm 100$ & \red{DA WD} & SOAR \\
3880049472115990400 &   &  20.34 & $-0.23 \pm 1.24$ & 87.0 & -1801 & 408 &  & \red{WB} &  \\
5160882249616496768 &   &  20.25 & $0.79 \pm 0.64$ & 121.7 & 730 & 404 & $-60\pm 50$ & \red{DA WD} & LRIS \\
6688913592127235584 &   &  19.19 & $0.41 \pm 0.29$ & 59.8 & 686 & 404 & $220 \pm 50$ & \red{DA WD} & SOAR/GMOS \\
3460197452350766464 &   &  20.48 & $-0.10 \pm 0.73$ & 52.2 & -2579 & 389 & $-20\pm 40$ & \red{DA WD} & SOAR \\
744576492507398272 &   &  20.30 & $-0.01 \pm 0.86$ & 65.9 & -56608 & 365 & $-180\pm 50$ & \red{DA WD} & LRIS \\
3792579576908228608 &   &  20.45 & $-0.28 \pm 1.08$ & 55.3 & -923 & 329 & $10\pm 50$ & \red{DA WD} & SOAR \\
3366889189061553664 &   &  20.47 & $-0.29 \pm 1.17$ & 56.3 & -906 & 303 & $-20\pm 50$ & \red{MS} & LRIS \\
4059697822386864512 &   &  20.49 & $-1.33 \pm 2.72$ & 85.7 & -306 & 290 & $0 \pm 50$ & \red{MS} & SOAR \\
4104739694138461184 &   &  20.23 & $-0.21 \pm 1.27$ & 56.5 & -1248 & 252 &  & \red{blended} &  \\
6217835085113310208 &   &  20.28 & $-0.14 \pm 1.19$ & 50.0 & -1739 & 225 & $-40\pm 50$ & \red{DA WD} & SOAR \\
5181766613992671104 &   &  20.49 & $-1.13 \pm 2.67$ & 56.6 & -236 & 174 & $80\pm 50$ & \red{DA WD} & LRIS \\
2552009991792563840 &   &  20.06 & $-2.01 \pm 0.96$ & 59.1 & -139 & -267 &   & \red{WB} &   \\
4096331041464263296 &   &  19.99 & $-1.42 \pm 0.82$ & 58.6 & -195 & -463 & $-120 \pm 50$ & \red{MS} & LRIS \\
5104365427803800832 &   &  20.18 & $-1.11 \pm 0.74$ & 56.9 & -243 & -742 & $-90 \pm 50$ & \red{DA WD} & LRIS \\
6278906633842892160 &   &  20.43 & $-1.00 \pm 0.85$ & 84.6 & -402 & -2676 & $70 \pm 50$ & \red{DA WD} & LRIS \\
6703717691563155968 &   &  19.24 & $-0.53 \pm 0.53$ & 52.7 & -471 & -79269 & $0\pm 50$ & \red{DA WD} & SOAR \\

\end{longtable}
\endgroup

\end{document}